**Household and Individual Economic Outcomes of Different Health Shocks: The Role of Medical Innovations**

**Volha Lazuka**

*Abstract*

This study provides new evidence on how medical care mitigates the economic consequences of health shocks for individuals and their partners. To identify causal effects, I focus on medical scientific discoveries and exploit longitudinal administrative data for Sweden, using a triple differences design. The results indicate that medical innovation strongly mitigates the negative economic consequences of health shocks for individuals and have spillover effects on their partners. These spillovers are relatively large because medical innovation compensates for partners' wage losses in conditions when welfare support for caregiving is insufficient. Overall, the findings indicate that medical innovation not only produces substantial economic gains but also reduces disease-related economic inequalities.

**Affiliations:** University of Southern Denmark, Lund University, and IZA. Address: University of Southern Denmark: Campusvej 55, DK-5230 Odense, Denmark. E-mail: vola@sam.sdu.dk.

**Acknowledgements:** I gratefully acknowledge funding support from the Jan Wallanders and Tom Hedelius foundation (grant no W18-0008) and Sweden's governmental agency for innovation systems, Vinnova (grant no 2014-06045). I wish to acknowledge Centre for Economic Demography (Lund University) for access to the individual-level data at Statistics Sweden. I thank the participants of the annual meeting of the Population Association of America, European Economic Association, COMPIE, and of the seminars at the University of Southern Denmark, Lund University, Linneaus University, and the University of Copenhagen. I wish to especially acknowledge the excellent assistance of Blaise Bayuo, Joe Bilsborough, and Alexander Generalov.

## I. Introduction

While the role of medical care in recovery from health shocks is well understood, little is known about its ability to mitigate the economic consequences of such shocks. Health shocks generate sizable economic losses for individuals (Frimmel et al. 2025; Fadlon and Nielsen 2021; Schmitz and Westphal 2017; García-Gómez et al. 2013), with the extent of these losses varying widely among individuals and often not being fully offset even when welfare transfers are considered (Meyer and Mok 2019; Lundborg, Nilsson, and Vikström 2015). The positive health effects of medical care across a wide range of diseases have become increasingly evident with the rapid and widespread advancement of medical technologies in recent decades (Murphy and Topel 2006). Several studies have established the economic impacts of specific medical innovations using experimental or quasi-experimental study designs.[1] However, the causal economic impact of the broad scope of medical innovations remains largely unstudied.

Health shocks impose substantial economic burdens that extend beyond the sick individual to their partners, who face difficult tradeoffs between maintaining earnings and providing informal care, and may ultimately become sick themselves (Jolly and Theodoropoulos 2023; Macchioni Giaquinto et al. 2022; Fadlon and Nielsen 2021; García-Gómez et al. 2013; Dobkin et al. 2018). These effects decline with more generous social insurance and greater access to formal care (Coyne et al. 2024; Anand, Dague, and Wagner 2022). However, medical innovation can benefit partners beyond welfare programs by reducing caregiving burdens and directly improving partners' deteriorated health. Medical innovation may thus improve economic welfare for entire families, yet the magnitude and distributional consequences of these gains have not been studied. Answering questions about the mitigating effects of medical innovation and its spillovers has important implications for the evaluation of the full economic returns to medical innovation and for the design of policies that support families facing health crises.

---

[1] The medical innovaions studied include pharmaceutical drugs and therapies for prostate and breast cancer (Jeon and Pohl 2019), sulfapyridine drugs against pneumonia (Lazuka 2020), pharmaceutical drugs and therapies for coronary heart disease (Stephens and Toohey 2022), antiretroviral therapy against AIDS (Thirumurthy, Zivin, and Goldstein 2008), Cox-2 inhibitors for arthropathies (Bütikofer and Skira 2018; Garthwaite 2012), and new technologies in hospital deliveries (Lazuka 2023).

Identifying the causal effects of health shocks and the mitigating effects of medical care on economic outcomes poses two key methodological challenges. The first challenge concerns isolating the causal impact of health shocks on economic outcomes. While some health shocks are sudden and largely unanticipated (e.g., cancers or heart attacks), most health shocks (e.g., mental, musculoskeletal, and a wide range of circulatory conditions) evolve through gradual health deterioration, even when they manifest as sudden events. As a result, these health shocks are often preceded by changes in behavior, employment, and health investments, which makes it difficult to disentangle the effect of the health shock itself from adaptive responses. Addressing these issues requires rich individual-level panel data and research designs that construct credible counterfactuals using variation in the timing of health events while validating pre-trends.

The second challenge concerns identifying and estimating the extent to which medical care mitigates the economic consequences of health shocks. The consumption of medical care is inherently endogenous, as it depends jointly on individuals' health status, expectations, and economic circumstances. It is possible to proxy medical care by medical innovation, such as the introduction of new drugs or therapies, though such innovation should capture meaningful variation across various conditions. But even when medical innovation is successfully measured at the disease level, identifying the causal mitigating effects of medical care on health shocks remains challenging, as appropriate counterfactuals must be constructed across both disease and medical-innovation dimensions. Finally, in the presence of potentially heterogeneous treatment effects across diseases, the chosen estimator should guarantee that estimated effects correspond to the target causal estimand without bias from negative weights.

This study applies a quasi-experimental design to examine health shocks across the full range of diseases observed in Swedish registers for adults aged 40–70 years. It leverages administrative register data spanning multiple cohorts across successive years and a constructed database of medical innovations available to treat 91 disease groups at the time of health shocks. The design is based on identifying, for each treated individual, a counterfactual individual who will be treated in the near future and is similar in pre-treatment observed characteristics. Based on the sample of treated and counterfactual individuals, the study first applies a difference-in-differences (DD) design and a two-way fixed-effects regression to estimate the impact of health shocks on the economic outcomes of both individuals and their partners. It then applies a difference-in-differences-in-differences (DDD) design and a three-way fixed-effects regression to estimate the mitigating impact of medical innovation on the outcomes, interpreted as an innovation-induced reduction in the economic losses caused by specific health shocks. The study further explores the mechanisms underlying these economic effects, including health and welfare outcomes for both individuals and their partners. Robustness checks assess both the validity of identifying assumptions and the sensitivity to alternative medical innovation measures.

The main formal assumptions underlying the design are: a parallel trends assumption—absent a health shock, treated and matched control individuals would have followed parallel trends in economic outcomes, both overall and at each level of disease-specific medical innovation; and a no-anticipation assumption—individuals do not anticipate either inpatient hospitalization or the arrival of pharmaceutical treatments. It is also assumed that compositional changes such as selective attrition due to mortality do not drive the results. Since fixed-effects regression is used as the main estimator, treatment effects are assumed to be homogeneous. Finally, while not critical for causal identification but important for accurately capturing magnitudes, the medical innovation measures should approximate the availability of medical care well.

The study uses data from the Swedish Interdisciplinary Panel, which provides longitudinal individual-level records linked through unique personal identifiers across registers. Health shocks are identified through inpatient hospital admissions, specifically first hospital admissions with no admissions in the prior three years and admissions with diagnoses aligning with medical innovation data pooled into 91 disease groups. I focus on individuals aged 40–70 and their partners, extracting information for 1978–2008, the widest period of overlap across the individual registers and time series of medical innovation. This yields 1,409,751 individuals who experienced a health shock at some point during this age range. The dataset provides rich information on economic outcomes, including real disposable family income and its components (earnings, capital income, sick-leave benefits, unemployment benefits, and early-retirement payments) for both individuals and their partners. To examine health mechanisms, I use outpatient hospital visits that approximate primary care visits.

For this study, I construct a database of medical innovation from official approval agencies in Sweden. The main medical innovation measure is based on a panel of new molecular entities (hereafter, pharmaceutical treatments) available to the patients through the healthcare system by disease group and year, compiled from Swedish Medical Products Agency records and validated against international sources. The panel includes 1,939 approved pharmaceutical treatments and 391 dissaproved pharmaceutical treatments. In the main analysis, I use a one-year lag of the cumulative count of pharmaceutical treatments, approved net of disapproved. An alternative measure—a cumulative count of granted net of lapsed patents in therapy, surgery, and diagnostics—is used in robustness analyses.

To address the first challenge of isolating the causal impact of health shocks, I employ a DD design. I first apply a matching strategy that constructs counterfactuals using variation in the timing of health events. Specifically, I match individuals who experience a health shock in a given year to observably similar individuals (on propensity score derived as a function of birth cohort, education, and earnings measured before age 40) who will experience the same health shock two years later. This combination of matching and shrinking the time window between hospitalized and not-yet-hospitalized individuals ensures that control individuals are on a similar health trajectory (with no detected influence from time-dependent unobservable factors) not only for severe and unanticipated diseases, but across the full range of diseases studied. Both a covariate balance test and test for the absence of differential pre-trends in economic outcomes between treated and control individuals validate that the design produces appropriate counterfactuals. As a result, I can obtain the causal effects of a health shock up to three years before and two years after the shock in an event-study and a DD specification.

The second challenge of identifying how medical care mitigates the economic consequences of health shocks is addressed by exploiting variation in pharmaceutical treatments both across diseases and over time in a DDD design. Pharmaceutical treatments exhibit substantial variation across the diverse conditions examined in this study from 1981 to 2006, with values ranging from zero to thirty. DDD specification uses variation in pharmaceutical treatments as a third dimension and isolates the treatment effect by controlling for both disease-specific trends and time-varying factors associated with different levels of pharmaceutical treatments. Because the control group consists of individuals who will experience the same disease in the near future, the design compares economic impacts of a health shock at different pharmaceutical treatment levels, yielding estimates that reflect the causal mitigating effect of medical innovation. The approach also avoids a bias from negative weighting and provides high-precision estimates.

This study has three main findings. First, an individual's health shock leads to substantial negative economic consequences for the family. The individual experiences a 5% income loss, but more strikingly, the partner suffers a 46% income loss. This disproportionate impact on partners arises because welfare transfers do not adequately compensate for partners' negative wage response, which is driven by the need to provide informal care and deterioration of partner's mental health. On the health dimension, a health shock increases outpatient hospital visits by 1.7 visits (nearly tripling the pre-shock mean of 0.6), reflecting the burden of ongoing care. This finding reveals that the economic burden of health shocks extends far beyond the affected individual, with partners bearing a particularly heavy cost in terms of income losses. Second, medical innovations substantially mitigate these negative consequences: one additional pharmaceutical treatment reduces family income loss by 1.7% (13% of the pre-shock mean) and reduces outpatient hospital visits by 0.03 (5% of the pre-shock mean). The beneficial effects of medical innovation are larger for the most vulnerable populations: individuals close to retirement age, those suffering from the most severe shocks, and those with lower pre-shock incomes. Third, one pharmaceutical treatment mitigates 10% of the initial income loss for individuals and 15% for partners. Partners accrue mitigation effects through two channels: through improvements in their own mental health due to medical innovation (10%) and through spillovers, as medical innovation reduces the partner's caregiving burden and enables greater labor force participation (5%). In sum, medical innovation generates substantial economic gains for both patients and their partners and reduces income inequality across individuals and within households.

This study makes two main contributions to the economics literature. First, it contributes to the applied microeconomic literature on the impact of medical innovations on economic outcomes by broadening the evidence to include all diseases observable in the population. Previous studies have examined how specific innovations—such as new drugs for particular cancers or treatments for cardiovascular diseases—affect labor market outcomes and earnings for affected individuals (Stephens and Toohey 2022; Lazuka 2020; Jeon and Pohl 2019; Bütikofer and Skira 2018; Garthwaite 2012; Thirumurthy, Zivin, and Goldstein 2008). This study shows that medical innovation's economic benefits are not limited to high-profile treatments for specific diseases but are widespread and substantial across the whole range of diseases. Furthermore, this broader range allows for assessment of heterogeneous effects across disease severity and patient vulnerability, demonstrating that innovations generate particularly large economic gains for the most disadvantaged populations. The findings also complement broader research on the aggregate productivity of medical care by introducing consumption of medical care as an important determinant of lifecycle income based on a quasi-experimental design (Cutler et al. 2021; Lakdawalla and Phelps 2021; Bloom et al. 2020; Lakdawalla, Malani, and Reif 2017; Papageorge 2016).

Second, this study contributes to the literature on spillovers of health shocks within a household. A robust body of economic literature demonstrates that adverse health events not only reduce the affected individual's earnings but also trigger household-level adjustments as spouses cope with these shocks by increasing their labor supply to compensate for lost household income or reducing work hours to provide caregiving (Jolly and Theodoropoulos 2023; Macchioni Giaquinto et al. 2022; Fadlon and Nielsen 2021; García-Gómez et al. 2013; Dobkin et al. 2018). These household-level economic consequences could be reduced by welfare family support (Coyne et al. 2024; Anand, Dague, and Wagner 2022) but are compounded by health spillovers, as caregiving partners frequently experience deteriorations in their own mental and physical health (Böckerman et al. 2025; Bom et al. 2019; Fadlon and Nielsen 2019). This study contributes to this literature by showing that health shock spillovers extend far beyond individual's income loss, as partners experience mental health deterioration and reduced labor force participation due to caregiving demands, with welfare transfers providing only partial (and small) compensation. This study contributes a novel dimension to the literature by demonstrating that medical innovation reduces adverse spillover effects within families through two mechanisms: improvements to partner's mental health (direct) and reductions in caregiving burdens through treatment of the individual's health (indirect). These mechanisms reveal that the economic value of medical innovation extends substantially beyond the effect on the individual to encompass broader household economic gains.

## II. Data

### A. INDIVIDUAL-LEVEL DATA ON INPATIENT HOSPITALIZATIONS AND OUTCOMES

I start with a description of the data to lay the foundation for the empirical strategy, which is described further. The data were then classified into the following two datasets: (1) data derived from income and health registers, which provide longitudinal individual-level records; (2) time series of pharmaceutical treatments for each disease taken from the databases of national approval authorities.

I obtained individual-level data from the Swedish Interdisciplinary Panel, which is a sample of the total Swedish population born between 1930 and 1985 and observed in administrative longitudinal registers.[2] The dataset includes data on health, demographics, and

[2] I used the database "Swedish Interdisciplinary Panel," which was hosted at the Centre for Economic Demography at Lund University (Statistics Sweden 1960-2021). Lazuka (2019) provides details about

socioeconomic outcomes linked through unique personal identifiers across registers. I included individuals aged 40–70 years, focusing on individuals in working age and around retirement. I extracted information on individuals and their partners for 1978–2008, the widest period of overlap allowed across the individual registers and time series of medical innovation.

To identify individuals who experienced health shocks due to specific diseases, I used inpatient hospital admissions data. Inpatient admissions—admissions with overnight stay—entail considerable economic consequences, are readily identifiable, and ensure access to the latest medical technologies including pharmaceutical drugs (Dobkin et al. 2018; Lundborg, Nilsson, and Vikström 2015). I applied four exclusion criteria. First, I restricted data to the first hospital admission of individuals with no admissions in the prior three years to minimize the chance of capturing anticipated shocks. Second, I limited admissions to cases for which a specific medical technology could be identified, excluding stays related to pregnancy, external causes, and symptoms. Third, the causes of hospitalization had to align with the data on pharmaceutical treatments. Finally, I restrict the sample to individuals who have a partner because of the focus on both the individual's and the partner's effects. These four criteria yielded 1,409,751 individuals who experienced a health shock at some point between ages 40 and 70 (the "treated" group).

The dataset provides a rich set of variables for determining outcomes. Since survival and inpatient hospitalization influenced treatment assignment, they could not be used as health outcomes. Therefore, I used the outpatient hospitalization register, which provides the number of visits per individual-year starting from 2001. Outpatient admissions—admissions without an overnight stay—take place in specialized care units that provide treatment not

---

the sources and reliability of the data.

available in primary care (Ludvigsson et al. 2025). Unlike inpatient admissions, outpatient admissions do not represent severe health shocks: they usually entail substantial waiting times and commonly refer to diagnostics (52% of all visits). Outpatient admissions typically follow a referral from a primary care doctor, so they may closely approximate primary care visits.

Regarding economic outcomes, I used real disposable family income as a main outcome variable because it has been regarded as the ultimate outcome of the economic consequences of a health shock (O'Donnell, van Doorslaer, and van Ourti 2015; Bolin, Jacobson, and Lindgren 2002). Moreover, it closely captures the efficiency implications of a health shock in settings with comprehensive public health insurance and negligible out-of-pocket expenses, as in Sweden. To study the mechanisms, I used personal disposable income and various economic variables that quantify its sources (earnings, capital income, sick-leave benefits, unemployment benefits, and early-retirement payments). Finally, I added information on the economic outcomes of the partner, including disposable income (from 1978) and its sources (from 1998).[3] Income information was included only for full calendar years when the individual was alive in order to avoid the influence of compositional changes across disease groups due to differential mortality. I transformed the economic outcomes using the inverse hyperbolic sine (IHS) to ease interpretation.[4]

[3] For the whole study period, 1978-2008, family income is defined in the registers as the sum of incomes for up to two co-resident generations who are related by marriage, cohabitation with a shared child(ren), or adoption and live at the same property. Partner income is calculated as family income minus the individual's personal income. For unmarried, non-cohabiting working-age adults living with their parents, this residual reflects parental—rather than partner—income. Starting in 1998, the data include unique partner identifiers, allowing me to directly link the partner's earnings, sick-leave benefits, and early-retirement payments.

### B. DATA AND MEASURES OF PHARMACEUTICAL TREATMENTS

Pharmaceutical treatments—the main measure of medical innovation in this study—are novel chemical compounds. Unlike pharmaceutical drugs, which can share the same compound but be marketed under different names (Kesselheim, Wang, and Avorn 2013), pharmaceutical treatments capture innovation at the compound level. The medical-innovation measure used here is a panel of pharmaceutical treatments available to the patients through the healthcare system by disease group and year, which I compiled in three steps.

First, I reached out to the Swedish Medical Products Agency to obtain a detailed register of all pharmaceutical drugs, their pharmaceutical treatments, and the dates of approval and disapproval for treating specific diseases in Sweden. Second, because each drug record provided the Anatomical Therapeutic Chemical (ATC) code of the underlying pharmaceutical treatment and its therapeutic indications, I matched each pharmaceutical treatment–indication combination to three-digit ICD codes using the Theriaque database (Husson 2008). These ICD codes are observed in the inpatient register. I then aggregated ICD codes into disease groups, balancing clinically meaningful categories—as defined by Elixhauser, Steiner, and

[4] IHS transformation may lead to the regression estimates being far from the true marginal effects when there are many zeros or negative values (Mullahy and Norton 2024) or when the values of the transformed variable are small (Aihounton and Henningsen 2021). In the estimation sample, across the main economic outcomes, the share of zero values reaches 4% at the maximum and there are no negative values. These outcomes, where units of measurement are typically large, were divided by 10,000 SEK before the transformation. Health outcomes contain a large share of zero values and are therefore not transformed.

Palmer (2015) —with the availability and consistency of hospitalization coding over the study period. The final list of 91 disease groups (see Appendix A, Table A1) was consulted with health experts (Lindström and Rosvall 2019). Finally, I validated the series by cross-checking the presence and timing of key pharmaceutical treatments against the World Health Organization Model List of Essential Medicines (WHO 2019).

The final database contains 6,743 pharmaceutical drugs and 1,939 pharmaceutical treatments, of which 391 were disapproved during the study period. I decided to keep pharmaceutical treatments that were eventually dissaproved because they also represent innovation for the period when they are permitted for practical use. In Sweden, pharmaceutical treatments undergo lengthy, stringent approval processes to ensure safety and efficacy; when a drug dissaproved, it may represent: (1) genuine safety withdrawals, (2) commercial decisions by manufacturers to cease marketing, (3) failure to reach pricing or reimbursement agreements with Swedish authorities, or (4) administrative deregistration for non-compliance with post-approval reporting requirements (Ljungkvist, Andersson, and Gunnarsson 1997). According to data, the most common reason for deregistration in Sweden is commercial rather than safety-related.[5] Consistent with the idea that most eventually disapproved pharmaceutical drugs are not cases of de-innovation, pharmaceutical treatments in the database that became later disapproved have an average survival time of 16 years.

My final measure of medical innovation is the cumulative count of pharmaceutical treatments available to the patients through the healthcare system (approved net of dissaproved) by year and disease group. I use this measure for three reasons: (1) a cumulative series captures the stock of medical knowledge; (2) in practice, combinations of new and older innovations are often most effective; and (3) a net measure helps mitigate autocorrelation within disease-specific series. The content and ranking of innovations generated by this series generally correspond to benchmark categorizations in the literature for pharmaceutical (Kesselheim and Avorn 2013).

[5] In the initial pharmaceutical drug list, only one pharmaceutical treatment —pentobarbital against kidney conditions—was deregistered in 1974 due to a high risk of fatal overdose.

Pharmaceutical treatments are ready for implementation in healthcare settings because each pharmaceutical treatment has a generic pharmaceutical drug name that is ready to be commercialized in the year of approval. To align with the timing of implementation, I use a one-year lag of pharmaceutical treatment: A recent analysis of drug availability finds that it takes about 281 days for a pharmaceutical drug to become available to Swedish patients after approval (IQVIA 2021). Most prior studies (Jeon and Pohl 2019; Lichtenberg 2015) choose the lag length after inspecting the empirical results, which makes formal hypothesis testing irrelevant. To facilitate comparisons with that literature, I also report results using longer lags in the robustness analysis.

Figure 1 illustrates the variation in pharmaceutical treatments underlying the DDD design across the sample period. The x-axis plots the initial stock of pharmaceutical treatments available for each disease in 1981, while the y-axis measures the growth in that stock through 2006. Each circle represents one or more disease groups, with its size proportional to the number of groups at that coordinate and the most significant diseases marked with shading. To see how this variation enters the DDD, consider two diseases. "Diseases of the gallbladder, biliary tract, and pancreas," near the bottom-left, started with roughly 2 treatments and experienced almost no pharmaceutical growth. "Ischaemic heart diseases," by contrast, began with a moderately higher stock and gained approximately 13 new pharmaceutical treatments over the same period. The DDD indicator is designed to capture two things: first, whether, within disease groups such as ischaemic heart diseases and gallbladder diseases, post-shock income losses are smaller when the stock of pharmaceutical treatments at the time of hospitalization is larger; and second, whether such disease-specific effects exhibit a linear relationship across levels of pharmaceutical treatments. A positively sloped fit is expected. For the case of heart and gallbladder diseases, this expectation—aligned with the literature on the ranking of medical innovations (Fermont et al. 2016)—implies that pharmaceutical treatments for heart diseases, compared to those for gallbladder diseases, were more efficient at the beginning of the period and that this relative efficiency increased to a larger extent over time. Figure A1 (Appendix A) shows the full variation in pharmaceutical treatments by year and disease group.

[Insert Figure 1 here]

## III. Empirical Strategy

### A. DD DESIGN FOR THE EFFECTS OF A HEALTH SHOCK

This study aims to define the extent to which medical innovations mitigate a health shock's negative consequences. This aim implies a causal inference; therefore, I applied a DDD approach and estimated medical innovations' impact on economic outcomes as an innovation-induced *reduction* in economic loss due to a health shock. This can be considered as the difference between the two DD estimators (Goodman-Bacon 2021). To form the first DD estimator ($treat_i \times post_{it}$), I compared the evolution of the economic outcomes of individuals between the treated and control individuals. I estimated the following equation:

$$(1)\ Y_{it} = \alpha_i + \beta_1 post_{it} + \beta_2 treat_i \times post_{it} + u_{it}$$

In this equation, $Y_{it}$ is an outcome for an individual $i$ in year $t$ . $treat_i$ denotes an indicator for whether an individual belongs to the treatment group; $post_{it}$ denotes an indicator for whether the observation belongs to the post-event periods. In our sample, $post_{it}$ is known for both treated and control individuals due to matching that eliminates the need to include year fixed effects (see Section III.B for details).[6] $\alpha_i$ represents individual fixed effects (which absorbs any time-invariant characteristics including the main effect of $treat_i$). An indicator $treat_i \times post_{it}$ (DD indicator) denotes post-event years (including the year of event) for treated individuals. For a health shock to reduce income, the estimate of $\beta_2$ should be negative and statistically significant.

## B. DDD DESIGN FOR THE MEDIATING EFFECTS OF PHARMACEUTICAL TREATMENTS

To form the second DD estimator, one needs to use the variation in $treat_i \times post_{it}$ by at least one more dimension; in our case, these differentially affected groups exist because the number of pharmaceutical treatments varies over time and across diseases. To obtain a DDD coefficient, where one of the differences varies across the values of a continuous variable (i.e., pharmaceutical treatments), I estimated the following DDD specification:

$$(2)\ Y_{itd} = \alpha_i + \beta_1 post_{it} + \beta_2 treat_i \times post_{it} + \beta_3 treat_i \times post_{it} \times M_{dt} + \\ + \lambda_1 M_{dt} \times post_{it} + \lambda_2 \alpha_d \times post_{it} + u_{itd}$$

In this equation, $M_{dt}$ represents one-year lagged pharmaceutical treatments available in year of hospitalisation $t$ to treat disease $d$ (see Section II.B for details). I will run several checks regarding pharmaceutical treatments in the robustness analysis.[7] The term $treat_i \times post_{it} \times M_{dt}$

[6] By contrast, in most staggered DD designs, $post_{it}$ is known only for the treated groups; therefore, year (time) fixed effects are necessary there.

[7] In Section E.III, I validate medical innovation measures through six robustness checks: (1) non-pharmaceutical innovations using Swedish patent data (30,687 patents across surgery, devices, diagnostics, and therapies); (2) first-difference transformations of pharmaceutical innovations; (3) exclusively international-origin pharmaceuticals; (4) five-year lags instead of one-year lags to account

denotes a DDD interaction of interest—it turns on for the treated individuals in the post-shock period at the level of $M_{dt}$.[8] For medical innovation to reduce the income loss from a health shock (i.e., increase income), the estimate of $\beta_3$ should be positive and statistically significant.

Eq.2 enables the exclusion of three main sources of bias from the main effect of interest $\beta_3$, which should represent the causal effect of a medical innovation on the outcome, i.e., the innovation-induced difference in the Average Treatment Effect on the Treated (ATT). First, the bias related to the permanent differences between individuals that affect both the outcome and treatment differs based on the presence of individual fixed effects $\alpha_i$.[9] For instance, if

---

for adoption time; (5) exclusion of drugs disapproved within five years; and (6) use of only pharmaceuticals that remained approved throughout the sample period. These validation tests are not critical for causal identification but important for accurately capturing magnitudes.

[8] In Eq.2, the effects of five terms—$treat_i$ (whether an individual is treated), $M_{dt}$ (one-year lagged pharmaceutical treatments available in year of hospitalisation $t$ to treat disease $d$), $\alpha_d$ (disease group-specific fixed effects), $treat_i \times M_{dt}$ (interactions of $M_{dt}$ with whether an individual is treated), and $\alpha_d \times treat_i$ (interactions of disease group-specific fixed effects with whether an individual is treated)—are absorbed by the individual fixed effects. This occurs because individuals receive separate identifiers when appearing in both the treated and control groups (making each final individual identifier unique), and for treated individuals, $M_{dt}$ is fixed at the level of disease group (corresponding to the main diagnosis) and year of hospitalization.

[9] As soon as an individual was matched, they received a new unique individual (experimental) number that was different from their original individual number. That is, observations for individuals who participated both as controls ($t \in [-8; -4]$) and then as treated ($t = 0$) are considered and constructed as being independent of each other.

higher-income or more educated individuals self-select into hospitalization for treating particular diseases, such (time-invariant) selection bias would be cancelled out in Eq.2. Second, changes in the outcomes over time—similar to all individuals—are also ruled out due to the inclusion of the post-event indicator $post_{it}$. Finally, it excluded time-varying bias specific to each level of pharmaceutical treatments, controlled by the interaction $post_{it} \times M_{ds}$, such as structural breaks in the post-event periods. The inclusion of the interaction between the disease groups and post-event dummies $\alpha_d \times post_{it}$ is necessary to complete a DDD specification.

### C. IDENTIFYING ASSUMPTIONS AND CONSTRUCTION OF THE COUNTERFACTUAL INDIVIDUALS

Assuming no anticipatory behavior, both DD and DDD designs rely on the parallel-trends assumption: there are no time-varying shocks specific to comparison groups—between treated and control groups in DD, and between treated and control groups at each level of pharmaceutical treatments in DDD (i.e., strong parallel trends assumption). Such assumption is a restriction on untreated potential outcomes trends that does not come from exogenous variation in inpatient hospitalization and in inpatient hospitalization at each level of pharmaceutical treatments (see related discussion in Baker et al. Forthcoming). Therefore, it is not guaranteed that such assumption is hold; for instance, if higher-income or more educated individuals seek treatment against particular diseases in a particular year when an effective pharmaceutical treatment becomes available, the resulting time-variying selection bias would violate the assumption.

To support the strong parallel trends assumption, I constructed the control group using an approach introduced by Fadlon and Nielsen (2021). They showed that individuals who would experience a heart attack or stroke in the near future serve as valid counterfactuals for individuals who experienced the same health shock in the year of analysis. I extend this approach to a broader set of diseases and conduct a formal *t*-test for the absence of pre-trends following Miller (2023). Table B1 in Appendix B presents the descriptive statistics of the final estimation sample.

I matched treated individuals to observably similar individuals who would experience the same health shock in the near future. Fadlon and Nielsen (2021) showed that valid counterfactuals can be obtained by matching individuals hospitalized or who died from these causes in year *t* to individuals hospitalized or who died from the same causes in year *t*+5. Because my study covers a broader set of diseases, I narrowed the window to *t*+2 and conducted matching within each of 91 disease groups.[10] Propensity scores were estimated using (i) year of birth (cohorts are widely dispersed), (ii) years of schooling, and (iii) real earnings (IHS) at ages 38–39, measured in the pre-event years since outcomes are observed from age 40 onward. Following Austin (2014), I used nearest-neighbor propensity score matching with a caliper of 0.2 standard deviations, without replacement.

From the original sample of treated individuals, I successfully matched 1,340,485 (95%) and then conducted two diagnostic checks on the matched sample. First, I tested for the covariate balance using absolute standardized differences with a threshold of 0.1, with values below commonly interpreted as indicating balance (Baker et al. Forthcoming; Austin 2009). All values fall below the threshold, as seen on Figure B1 in Appendix B that reports these diagnostics for the full sample and by ICD chapter. Because covariate balance alone does not guarantee parallel trends in a DDD design, I also examined mean economic outcomes by comparison group across event years—before and after the health shock.

[10] This is the smallest window possible: For the pre-treatment period, three years is the minimum time to detect non-linearity in outcomes based on *t* and *F*-tests; for the treatment period, the year after hospitalization, *t+1*, is the first year when the negative effect of hospitalization is fully realized.

To further investigate pre-trends in a DD design, Figure 2 plots the means of the main economic outcomes—family income, personal income, and partner income—for the treated and control groups across event years. Appendix B, Figures B2–B4, show analogous graphs disaggregated by disease group and convey a similar picture. The outcomes evolve similarly prior to event time $t = 0$ (the hospitalization), consistent with the matching procedure working as intended. At and after $t = 0$, family income declines sharply among treated individuals, indicating the onset of economic loss at the family level, whereas the control group shows no comparable change.

[Insert Figure 2 here]

To detect pre-trends in the DDD design, I conducted two formal tests to assess the absence of nonlinear pre-trends in family income, separately by disease group.[11] First, following Borusyak, Jaravel, and Spiess (2024), I estimated a fully dynamic (event-study) version of the underlying DD models, treating $t = -3$ and $t = -1$ as reference categories and using an $F$-test to detect nonlinear pre-trends. I estimated these models for each of the 91 disease groups, separately for men and women. However, because this test is sensitive to sample size, it tends to indicate the presence of pre-trends even when they are economically negligible. To address this, Rosenbaum and Rubin (1985) recommend the standardized difference, which is insensitive to sample size. Accordingly, as a second test, I calculated standardized differences in outcomes between treated and control individuals for each disease group.

[11] The pre-trends test is conducted across 91 disease groups, while the strong parallel trends assumption requires that there are no time-varying shocks between treated and control groups at each level of pharmaceutical treatments. However, it is highly plausible that the test validates the strong parallel trends assumption. First, pharmaceutical treatments in the sample vary between 0 and 30, spanning fewer distinct innovation levels than there are disease groups. Second, even though pharmaceutical treatments vary within each disease group over time, they do not differ drastically between comparison groups because the construction of counterfactuals is performed across closely matched individuals (those hospitalized in year $t$ versus year $t+2$ due to the same disease group).

Most disease groups passed both tests (see Table B2 in Appendix B). Of 91 groups, 89 showed no pre-trends at the 5% level in the *F*-test. One exception—in situ neoplasms—exhibited statistically and economically meaningful pre-trends. For ischemic heart disease, the *F*-test indicated a 0.6% pre-event income gap, while the health shock reduced income by almost 60%, indicating that such small pre-trends could not nullify the shock's impact. In the standardized-difference test, 88 groups were below the 0.1 threshold; the remaining three showed only marginal imbalance. Together, these tests support similar pre-event development of the outcomes between treated and control individuals by a disease group. An earlier version of this study (Lazuka 2021), which excluded several disease groups with significant pre-trends, yielded estimates for the full sample nearly identical to those reported in the present study. Therefore, given the similarity of results and the study's focus on broad disease coverage, I retain the full sample of 91 disease groups.

## IV. Main Results

### A. DD RESULTS: EFFECT OF A HEALTH SHOCK ON FAMILY INCOME

I begin by examining the effect of a health shock on family income. Conceptually, family income is the ultimate outcome of the economic consequences of a health shock, and its overall response is ambiguous because it reflects multiple channels—typically negative for the affected individual and uncertain for other household members (O'Donnell, van Doorslaer, and van Ourti 2015; Bolin, Jacobson, and Lindgren 2002). In generous welfare settings, family income often declines following an individual's health shock (Fadlon and Nielsen 2021; García-Gómez et al. 2013; Riphahn 1999).

Table 1 presents the estimated effect of a health shock on family income—namely, DD coefficient denoted by $\beta_2$ in equation (1), over two years (column 1) and separately by event year (column 2). Consistent with previous studies, the results show that families experience a net income loss when an individual suffers a health shock. On average, family income declines by 31%. The loss narrows slightly in the second year—from 33% to 30%—but remains substantial. All effects are precisely estimated and statistically significant at the 1% level. In absolute terms, given an average annual family income of SEK 320,803 (USD 33,466), the loss amounts to SEK 103,331 (USD 10,779) per person-year. The effect size is comparable to the earlier results in Figure 1, which show the trajectory of average family income for the treatment and control group.

[Insert Table 1 here]

### B. DDD RESULTS: MEDIATING EFFECT OF PHARMACEUTICAL TREATMENTS ON FAMILY INCOME

In this section, I examine how medical innovation mediates the effect of health shocks on family income. Medical care—approximated here by medical innovation—is a key input in family health production; knowing its returns (in terms of economic responses) is relevant for both households and policymakers. Moreover, if medical care mitigates the adverse consequences of a health shock, then failing to account for differences in medical care will underestimate the full economic impact of such shocks. The marginal value of medical innovation may increase with disease severity due to both consumption gains (direct utility from health) and income gains (productive time) (Grossman 2000). Empirical evidence has been provided for the consumption mechanism (Lakdawalla and Phelps 2020), but evidence for income gains from medical innovation, especially with disease severity, remains scarce.

I first estimate the effect of a health shock on family income across medical innovation quartiles —this allow to show how the effect of medical innovation emerges. Specifically, I estimate a fully-dynamic (event-study) specification of equation (1) on four subsamples defined by $M_{dt}$: lowest quartile (0–3 pharmaceutical treatments), second quartile (4–7 pharmaceutical treatments), third quartile (8–14 pharmaceutical treatments), and highest quartile (15–30 pharmaceutical treatments). This setup yields event-study (DD) coefficients that are interpretable as ATTs because they compare 0 to each level of medical innovation, owing to the sample construction: treated individuals (diagnosed with disease *d* in year *t* and thus exposed to $M_{dt}$) are matched to individuals observed in year *t* who will be treated for disease *d* two years later and are therefore exposed to $M_{dt} = 0$ (a related discussion is provided in Callaway, Goodman-Bacon, and Sant'Anna 2024).

Figure 3 presents event-study estimates of health shock effects on family income across medical innovation quartiles. The results reveal a declining impact as pharmaceutical treatment availability increases: in the lowest quartile (0–3 pharmaceutical treatments), income losses reach 43% in both the shock year and the following year. In the highest quartile (15–30 pharmaceutical treatments), losses fall to 16% in the shock year and 13% the following year. The estimates are precise with narrow confidence intervals, confirming that the reduction in health shock impact is statistically significant across levels of pharmaceutical treatments. Critically for identification, pre-shock estimates ($t = -2$) are not statistically different from zero, indicating no differential pre-trends across medical innovation levels. Using the median points of the lowest and highest quartiles (2 and 18 pharmaceutical treatments, respectively), the reduction in income loss between these quartiles is 27 percentage points (43% − 16%) in the shock year. This implies approximately 1.7 percentage points of loss reduction per additional pharmaceutical treatment [27 / (18 − 2)].

[Insert Figure 3 here]

Next, I estimate the mediating effect of medical innovation on family income, captured by the DDD coefficient $\beta_3$ in equation (2). Table 2 reports estimates for the two years following the health shock (column 1) and separately by event year (column 2). The results indicate that one pharmaceutical treatment increases family income by 1.7%, an effect that is statistically significant at the 1% level. This effect aligns closely with the reduction in shock-related income loss per additional pharmaceutical treatment shown in Figure 2. Across event years, the effect of medical innovation increases modestly—from 1.6% to 1.8%—consistent with innovation reaching its full impact in a year since the shock. In absolute terms, one pharmaceutical treatment increases family income by 5,133 SEK (USD 536) per person-year.

[Insert Table 2 here]

The mediating effect of medical innovations—a 1.7% increase in family income per unit increase in pharmaceutical treatment—suggests their high economic efficiency. First, it is useful to juxtapose this result with estimates of the aggregate productivity of medical care. Recent studies that consider realized utilization of medical care and labor-productivity growth estimate annual productivity of medical care for the working-age population at about 0.7% (Fonseca et al. 2021), which is roughly 2.5 times smaller. The effect estimated here is compatible because, in the Swedish setting, family disposable income is measured net of taxes and thus already excludes medical expenses. As I show below, the main reason for the difference is spillover effects not accounted for in previous studies. Second, I relate the effect to the gap in responses to a health shock between poorer and richer individuals. In additional analyses (available upon request), I find that individuals who have financial debts experience a 34.3% reduction in family income due to a health shock, versus a 27.4% reduction for the individuals without debts. To close the 6.9 percentage-point gap between these groups (34.3 − 27.4), society would need to introduce four pharmaceutical treatments (6.9 ≈ 1.7 × 4); this appears feasible, as one standard deviation in pharmaceutical treatments is seven units.

### C. DD AND DDD RESULTS FOR PERSONAL AND PARTNER INCOME

In this section, I will examine the impact of a health shock (DD) and the mediating effect of medical innovation (DDD) on family members—individuals and their partners.

Table 3 presents the effects of the health shock on personal income (column 1) and partner income (column 2), reporting DD coefficients over two years and separately by event year. The results show that individuals, due to a health shock, experience a 5% reduction in personal income. It is plausibly a permanent deterioration in health capital, as the shock-induced reduction in income does not diminish over time. In the European context, partners and children often reduce labor-force participation to provide informal care and to compensate for the individual's reduced household productivity (Frimmel et al. 2025; García-Gómez et al. 2013). Consistent with this evidence, partner incomes decline by 46% following an individual's health shock that is substantially larger than the losses for the affected individuals themselves. These changes also appear permanent, as the gap in this outcome persists into the second year after the health shock. In absolute terms, an individual's health shock reduces personal income by 9,455 SEK (USD 987) and partner income by 64,329 SEK (USD 6,717) per person-year.

[Insert Table 3 about here]

To visualize the mediating effect of medical innovation, Figure 4 presents event-study estimates of health shock effects on personal income (Panel A) and partner income (Panel B) across pharmaceutical treatment quartiles. For both individuals and their partners, the impact of a health shock declines with increasing pharmaceutical treatment availability. Moving from the lowest quartile (0–3 pharmaceutical treatments) to the highest quartile (15–30 pharmaceutical treatments), personal income losses fall from 10–11% to approximately 0%, while partner income losses decline from 58–62% to 20–25%. These estimates imply a loss reduction of approximately 0.7 percentage points per additional pharmaceutical treatment for personal income [(10% − 0%) / 16 pharmaceutical treatments] and 2.4 percentage points per additional pharmaceutical treatment for partner income [(62% − 24%) / 16 pharmaceutical treatments]. As with family income, pre-shock event-study estimates for both personal and partner incomes are not statistically different from zero, indicating no violation of the parallel-trends assumption.

[Insert Figure 4 here]

Proceeding to Table 4 (columns 1 and 2), I present DDD estimates of the mediating effects of medical innovation on personal and partner income. Consistent with Figure 3, one additional pharmaceutical treatment leads to a 0.5% increase in personal income and a 2.1% increase in partner income; both effects are statistically significant at the 1% level. Across event years, the effect on partner income rises from 1.9% to 2.3%, while the effect on personal income remains stable. In absolute terms, one additional pharmaceutical treatment generates SEK 913 (USD 95) higher personal income and SEK 2,946 (USD 308) higher partner income per person-year.

[Insert Table 4 about here]

Although the percentage effects are larger for partners than for individuals, they represent smaller shares of the initial health shock that medical innovation mitigates: 10% (0.5% / 5%) for personal income versus only 4% (2% / 50%) for partner income. I will explore the reasons why medical innovation mitigates a smaller share of partner income loss in the next section.

### *D. Mechanisms*

#### D.I. DD AND DDD RESULTS FOR OUTPATIENT HOSPITAL VISITS AND THE SOURCES OF INCOMES

In this section, I study the mechanisms of income effects of a health shock and medical innovation by examining health and the sources of personal and partner income as outcomes, and the heterogeneity of income effects across age and disease dimensions.

The estimates for outpatient hospital visits of the individual are presented in Table 5 (column 1). Results indicate that a health shock increases the number of hospital visits by 1.7 visits (or by 2.8 times the pre-mean of 0.6). One additional pharmaceutical treatment reduces hospital visits by 0.03 (about 5% of the pre-mean). This effect is substantial: increasing the number of pharmaceutical treatments across the median points in the lowest and highest quartile of pharmaceutical treatments distribution—from 2 to 18—reduces hospital visits by 0.5, offsetting one third of the increase caused by a health shock.

[Insert Table 5 about here]

Regarding the sources of income loss due to a health shock (see Table 5, columns 2-7), the reduction in wages by 38% (83,008 SEK [USD 8,636]) among the treated individuals is compensated by the receipt of various welfare benefits. Specifically, in relative terms, there is a 2.4-fold increase in sickness absence benefits (a job-based income insurance covering periods of short-term sickness), a 33% increase in unemployment payments, an 18% increase in disability payments, an increase in early retirement (6%) and self-insurance through capital income (4%). The mediating effects of medical innovation reduce both wage losses (increasing wages by 1.6% per pharmaceutical treatment) and reliance on major welfare benefits due to restored health capital (–0.2% for sickness benefits, –0.5% for unemployment, –0.3% for disability, and –0.4% for early retirement per pharmaceutical treatment).

For the partner, I present the estimates for the outpatient hospital visits due to mental diseases and the sources of partner incomes as outcomes in Table 6. While partner incomes are available for the period 1978–2006, the sources of partner incomes can only be observed after 1998. As before, the results show a substantial decrease in partner income due to an individual health shock: a 23% reduction (19,274 SEK [USD 2,005]), which is smaller than the reduction for the full sample (46% in Table 3), as income losses diminish over time.

[Insert Table 6 about here]

Regarding the sources of income decline, they relate to both partners' mental health and employment. As a result of the individual's health shock, partners' outpatient hospital visits for mental diseases (primarily anxiety and depression) increase by 0.004 visits, or 14% of the pre-treatment mean. There are no effects for other diseases. The effects for wages also show a decline by 14%. Financial compensation for time off to care for a seriously ill close relative was introduced in Sweden in 2003 (Justitiedepartementet 2003) and is captured in sickness or disability benefits. Indeed, compensatory effects appear in both sickness benefits (7%) and disability benefits (2%). The larger effect for sickness benefits reflects both partner compensation for caregiving and compensation for deterioration of their own (mental) health.

Turning to the mitigating effects, one pharmaceutical treatment increases partner income by 1.3%, offsetting 5% of the initial income loss—similar to the relative reduction observed over the full study period. This smaller mitigation for partners (5%) compared to individuals (10%) reflects that mediating effects of medical innovation appear only in wages (0.8% increase, or 5% of the initial wage shock) but not in outpatient hospital visits for mental diseases. This suggests that partners' mental health deterioration occurs at the extensive

margin—triggered by the individual's health shock—serving as a useful placebo test since medical innovation targets the individual's diagnosis, not the partner's mental health. However, partners also have access to medical care; hence, medical innovation should impact them directly, through improvements in their own mental health. I could infer the beneficial causal effects of mental health care from the individual's estimates and thereby estimate the effects of health shocks and of medical innovation after 2001 on a subsample of mental health diagnoses (see Table C1, Appendix C). These estimates show that one pharmaceutical treatment reduces income loss by 0.7%, primarily by reducing doctor visits and receipt of sickness benefits by 10% each relative to baseline. Taken together, medical innovation mitigates 15% of the income loss for partners (10% directly from medical innovation and 5% indirectly through spillovers).

Effects of medical innovation on sickness and disability benefits for the partners are marginally stiatistically significant and become more precise after 2003, when payments for caregiving relatives were introduced. In sum, medical innovation mitigates the negative economic consequences of health shock for individuals and for partners when sufficient welfare support is absent.

### D.II. HETEROGENEITY

To further explore the mechanisms, I examine the presence of heterogeneous mediating effects by age, disease severity, and pre-shock income and present the results for family, personal, and partner income in Table 7.

[Insert Table 7 about here]

The mediating economic impact is ambigious in relation to age. On the one hand, more efficient medical innovations tend to be developed for diseases that are more common in older age (Cutler et al. 2021). On the other hand, many older individuals are retired, whereas medical innovations primarily help restore health and enable non-retired individuals to return to work (Galama et al. 2013). The results show that family income increases by 1.2% per pharmaceutical treatment (SEK 1,174 [USD 122]) for younger individuals, by 4.2% per pharmaceutical treatment (SEK 1,394 [USD 145]) for older non-retired individuals, and by 0.3% per pharmaceutical treatment (SEK 84 [USD 9]) for older retired individuals. Consistent with our earlier finding that medical innovation reduces early retirement, the larger effects for older but non-retired individuals suggest that earlier care helps them return to work sooner and discourages them from exiting the labor market.

Regarding disease severity, previous literature has established that treating more severe diseases yields greater economic gains, as patients place higher value on treatment for severe diseases (Lakdawalla and Phelps 2020). In applied economics literature, large earnings effects from medical innovation have also been documented for breast and prostate cancer patients (Jeon and Pohl 2019). The results of the present study show that gains from medical innovation in terms of family income are about 10 times larger for cancer than for other diseases: a 5% increase per pharmaceutical treatment (or SEK 1711 [USD 178]) compared to a 0.5% (or SEK 146 [USD 15]) increase per pharmaceutical treatment (columns 3 and 4).

Given that medical innovation exhibits larger mitigating effects for older and sicker individuals, I examine its distributional impacts more generally—by estimating effects across pre-shock income quartiles. Table C2 in Appendix C reveals that the mitigating impact of pharmaceutical treatments decreases with income level, with the smallest effects concentrated in the highest income quartile. For family income, one additional pharmaceutical treatment increases income by 1.9% in the lowest quartile compared to 1.3% in the highest quartile.

Finally, the mitigating effects of medical innovation are consistently larger for partner income than for personal income among individuals above age 60 or those diagnosed with cancer—groups that require more caretaking when efficient medical innovations are absent.

Since the measure of medical innovation is continuous (comparing the effects of moving from fewer to more pharmaceutical treatments; see Figure 2), these results suggest that newer innovations are more effective than older ones to mitigate the burden of additional care on the part of the partners. This efficiency supposedly stems from the ability of new pharmaceutical treatments to mitigate disease symptoms (as observed in sickness benefits as an individual outcome; see Table 5) and/or to reduce the amount and severity of side effects.[12]

In sum, medical innovation not only generates substantial economic gains but also reduces economic inequalities both among sick individuals and between sick individuals and their partners, who are otherwise affected by the need to provide additional care without being fully compensated.

### *E. Robustness Analyses*

In this section, I present the results of various robustness analyses. As shown below, these analyses do not alter the main findings of the paper.

#### E.I ACCOUNTING FOR COMPOSITIONAL CHANGES

In the case of unbalanced panel data, compositional changes due to time-varying selective attrition (selective entry or exit) may bias the estimated ATTs (Sant'Anna and Xu 2023). In the present study, selective entry is not an issue as I restrict the sample to individuals observed for at least three years before treatment; imbalance may arise from selective exit due to differential mortality, outmigration, or divorce across treatment groups.

[12] The newer cancer drugs and technologies are generally more comfortable to the patient and less toxic, with fewer severe and life-threatening conditions requiring care and monitoring (Kroschinsky et al. 2017). New anti-cancer therapies result in better treatment outcomes and fewer side effects (Bradley 2008).

Additionally, specifically to the DDD estimator, a bias may be induced by the compositional differences between treatment groups for different cohorts (in our case, levels of medical innovation), as suggested in Goodman-Bacon (2021). Even though the results in this paper are similar across event years suggesting the absence of bias (see, for instance, Table 2 column 2), I conduct several robustness checks.

I present the results of the robustness checks accounting for compositional changes in Appendix D, Table D.1. A straightforward way to assess whether compositional changes bias the results is to restrict the estimation sample to individuals with balanced panel observations over five years that results in a loss of only 4% initial observations (see Panel I, Table D.1). As it might be important to keep all individuals to obtain population-level ATTs, I alternatively regress the probability of being observed in only one event-year after the shock on whether the individual has financial debts in a year before, predict probabilities, and then reweight the main regression estimates using inverse probability weights (see results in Panel II, Table D.1). As a final check, to account for compositional differences between treatment groups in the DDD estimator, I add event-year fixed effects interacted with ICD-chapter disease groups to equation (2) (see Panel III, Table D.1). Across the three checks, the results are similar or identical to the main results of the paper, suggesting that compositional changes do not introduce bias in this case.

### E.II AN ALTERNATIVE ESTIMATOR

Modern methodological literature has revealed that OLS regressions with fixed effects may produce estimates far from ATT in the presence of heterogeneous effects, due to a weighting problem (Callaway, Goodman-Bacon, and Sant'Anna 2024; Callaway and Sant'Anna 2021). The solution proposed in this literature to solve the problem—estimating the cohort-average treatment effects and appropriately aggregating them—is similar to the empirical approach applied in the present study.[13] As mentioned earlier, I matched each treated individual to the not-yet-treated individual, extracted the same pre- and post-treatment years for each pair, and stacked all pairs with duplicates in regressions. This solved two problems related to weighting. First, there were no negative weights in my estimation, meaning that the DD and DDD estimates could not be of different signs compared to the ATT. Second, the availability of treatment pairs observed for the same period ensured that differential treatment groups received equal weights and contributed equally to the estimates in the two- and three-way fixed-effects regressions.

As a robustness check, I use Callaway and Sant'Anna's (2021) estimator to estimate the mitigating effects of medical innovation on incomes for the lowest and highest quartiles of the distribution of medical innovation (see the estimates in Appendix D, Table D.2). The results show that the reduction in family income is 40% when pharmaceutical treatments fall into the lowest quartile and 9.8% when they fall into the highest quartile. Using the median points of the lowest and highest quartiles (2 and 18 pharmaceutical treatments, respectively), the reduction in income loss between these quartiles is 30 percentage points in the shock year. This implies approximately 1.9 percentage points of loss reduction per additional pharmaceutical treatment [30 / (18 − 2)]. For personal and partner income, the corresponding reductions amount to 0.8 and 1.7 percentage points. These estimates are not statistically different from the main estimates.

[13] The methodological DD literature has proposed several approaches to estimate ATT effects for the case of staggered treatment timing that are robust to heterogeneity over time and space. However, Callaway and Sant'Anna's group-time estimator is a generalization of other popular approaches (Roth et al. 2023).

### E.III ALTERNATIVE MEASURES OF MEDICAL INNOVATION

Pharmaceutical treatments had advantages over other medical innovation measures (see Section II.B for details), but other measures may be preferred, as shown in this subsection. I present the DDD estimates using alternative measures of medical innovation in Table D.3 Appendix D.

First, I use indicators of non-pharmaceutical medical innovations (see Panel I in Table D.3, Appendix D). I construct them using patent data from the Swedish Patent Database run by the Swedish Patents and Registration Agency. I limited the International Patent Classification codes to those covering surgery, electrotherapy, magnetotherapy, radiation therapy, ultrasound therapy, medical devices, and diagnostics.[14] Then, based on the names of diseases in the corresponding ICD versions within each disease group, I formulated combinations of keywords to conduct inclusive yet independent searches (available upon request). Based on the IPC codes and keywords, I conducted a search for the number of patents granted and lapsed per disease group and year in the heading and text of patents. My final database contains 30 687 granted patents, out of which 3 921 were lapsed.

Second, I use different transformations of pharmaceutical medical innovations. Prior literature has used pharmaceutical innovations of exclusively international origin or flow measures such as first differences, arguing these specifications enhance exogeneity (Lichtenberg 2019; Papageorgiou, Savvides, and Zachariadis 2007). While my main identification strategy does not rely on the exogeneity assumption of medical innovations but instead on the strong parallel trends assumption, imported innovations and newest innovations (as measured by first differences) may be most economically significant. Therefore, I rerun models with first differences of medical innovations (see Panel II) as well as medical innovations of exclusively international origin (see Panel III). Furthermore, one-year lags of pharmaceutical treatments may not adequately capture the medical innovations used in practice, since it can take time for pharmaceutical treatments to be adopted. I therefore estimated the models with five-year lags (see Panel IV).  Finally, I conduct two robustness checks validating the inclusion of eventually disapproved pharmaceutical treatments in the medical innovation measure: 1) I exclude pharmaceutical treatments disapproved within five years of approval, 2) I use only pharmaceutical treatments that remained approved throughout the sample period (see Panels V and VI).

---

[14] They correspond to the categories linked to diagnostics, therapy and surgery in subchapter in A61 "Medical or Veterinary Science; Hygiene." I excluded patents granted for A61K "Preparations for medical, dental, or toilet purposes," which makes the variable measuring patents complementary to that for pharmaceutical drug approvals.

The results of the robustness checks show mitigating impacts of medical innovation of similar magnitudes to the main results of the paper. When patents (in tens) are used as a measure, medical innovation increases family income by 3.5%; in terms of one standard deviation, this impact equals 1.5% (3.5% / 2.4), while the impact of pharmaceutical treatments in terms of one standard deviation equals 2% (1.7% / 0.8). Pharmaceutical treatments and patent-based measures of medical innovation correlate at 33%, suggesting that the pharmaceutical treatment measure also captures non-pharmaceutical innovation in mitigating the economic impacts of shocks. Excluding eventually disapproved pharmaceutical treatments reduces the estimated effect from 1.7% to 1.4%, indicating that disapproved pharmaceutical treatments provide meaningful economic benefits and should be included in the measure when they are permitted for practical use (as in the main specification). Other specifications generate somewhat larger effect sizes, consistent with a priori expectations.

## V. Conclusions

Despite growing evidence of the negative economic consequences of various health shocks and their heterogeneity in different treatment schemes, little is known about the extent to which these consequences can be mitigated by medical care. This study fills this gap in the literature by studying adults in Sweden aged from 40–70 years suffering with diseases of varying severity and progression and spillovers to their partners. To obtain the causal effects of medical care, I focused on the role of medical scientific discoveries and leveraged the longitudinal dimension of administrative microdata.

This study reveals that medical innovations have sizable mitigating effects on the economic outcomes of individuals and their partners, and that these effects reduce income inequalities caused by health shocks. One additional pharmaceutical treatment reduces family income loss (and therefore increases family income) by 1.7%, or USD 536 per person-year (in 2021 dollars). The mitigating economic effects of medical innovations are larger for partners than for individuals (2.1% or USD 308 versus 0.5% or USD 95), as well as for the most vulnerable populations, such as older non-retired individuals (4.2% versus 1.2% for younger individuals) and those suffering from cancer (5% or USD 178 versus 0.5% or USD 15 for other diseases). Efficient medical care reduces not only welfare transfers intended to compensate for the individual's capacity and income losses (by 0.2–0.5% per pharmaceutical treatment) but also financial compensation for time off to care for a seriously ill close relative (by marginally significant amounts following the 2003 introduction of this benefit).

This study provides direct policy implications. First, it shows that medical innovations can be regarded as investments with high returns and provides the estimated per-pharmaceutical treatment effects for the household, individual, and partner. Policymakers can directly translate these effects into cost-benefit calculations for R&D subsidies and apply distributional weights based on the effects across subgroups to capture equity objectives (Wild et al. 2025). Second, the study finds that medical innovation directly reduces welfare transfers for individuals and increases wages for partners. In drug pricing and approval decisions, these estimates can be incorporated into extended cost-effectiveness analyses, valuing these benefits alongside direct health gains (Hult and Philipson 2023). Finally, causal estimates of economic losses due to health shocks, along with the mitigating effects of medical innovation, can guide safety-net design by showing how insurance premiums could be adjusted to account not only for the reduced health risks from medical innovation (as in Lakdawalla, Malani, and Reif 2017) but also for the reduced economic risks faced by patients and their partners.

Table 1. Impact of a Health Shock on IHS *Family Income*: Baseline DD Specification

| | Family income | Family income |
|---|---|---|
| | (1) | (2) |
| $treat_i \times post_{it}$ | -0.313*** | |
| | (0.002) | |
| by event year | | |
| | | |
| $treat_i \times$ *event year 0* | | -0.326*** |
| | | (0.002) |
| $treat_i \times$ *event year 1* | | -0.300*** |
| | | (0.002) |
| | | |
| Outcome for $treat_i \times post_{it} = 0$ | 32.803 | 32.803 |
| Observations | 11 032 884 | 11 032 884 |
| Number of individuals | 2 243 040 | 2 243 040 |

*Notes*: This table reports estimates of the regression coefficient on $treat_i \times post_{it}$ from equation (1) (column 1), and from an event-study specification that replaces $treat_i \times post_{it}$ with interactions for event-year 0 and event-year 1 (column 2). The sample includes treated individuals—who experience a health shock (inpatient hospitalization) —and control individuals who experience the same health shock three years later. Each treated inidvidual is matched to the control individual within the same sex, disease group (91 groups), and year of hospitalization, as well as on the propensity score constructed from pre-event characteristics: year of birth, years of schooling, and earnings (IHS) (see Section III.B for details). Outcome for $treat_i \times post_{it} = 0$ is provided for the absolute outcome in 10,000 SEK. Individuals receive separate identifiers when they appear in both the treated and control groups. Standard errors, clustered at the individual (experimental) level, are reported in parentheses.
*** $p < 0.01$, ** $p < 0.05$, * $p < 0.10$

Table 2. Mediating Impact of Pharmaceutical Treatments on IHS *Family Income*: DDD Specification

| | Family income | Family income |
|---|---|---|
| | (1) | (2) |
| $treat_i \times post_{it} \times M_{dt}$ | 0.0165*** | |
| | (0.000) | |
| by event year | | |
| $treat_i \times M_{dt} \times \textit{event year 0}$ | | 0.0155*** |
| | | (0.000) |
| $treat_i \times M_{dt} \times \textit{event year 1}$ | | 0.0175*** |
| | | (0.000) |
| | | |
| Outcome for $treat_i \times post_{it} = 0$ | 32.803 | 32.803 |
| Observations | 11 032 884 | 11 032 884 |
| Number of individuals | 2 243 040 | 2 243 040 |

*Notes*: This table reports estimates of the regression coefficient on $treat_i \times post_{it} \times M_{dt}$ from equation (2) (column 1), and from an event-study specification that replaces $treat_i \times post_{it} \times M_{dt}$ with interactions for event-year 0 and event-year 1 (column 2). The sample is the same as in Table 1. $M_{dt}$ denote pharmaceutical treatments that are a one-year lag of cumulative count of new molecular entities available in year $t$ to treat disease $d$ (see Section II.B for details). For treated individuals, $M_{dt}$ are fixed at the level of disease group and year of hospitalization. Individuals receive separate identifiers when they appear in both the treated and control groups. Outcome for $treat_i \times post_{it} = 0$ is provided for the absolute outcome in 10,000 SEK. Standard errors, clustered at the individual (experimental) level, are reported in parentheses.

*** $p < 0.01$, ** $p < 0.05$, * $p < 0.10$

Table 3. Impact of a Health Shock on IHS *Incomes of Family Members*: Baseline DD Specification

| | Personal income | Partner income |
|---|---|---|
| | (1) | (2) |
| $treat_i \times post_{it}$ | -0.050*** | -0.463*** |
| | (0.002) | (0.004) |
| by event year | | |
| | -0.053*** | -0.475*** |
| $treat_i \times event\ year\ 0$ | (0.002) | (0.005) |
| | -0.048*** | -0.452*** |
| $treat_i \times event\ year\ 1$ | (0.002) | (0.005) |
| | | |
| Outcome for $treat_i \times post_{it} = 0$ | 18.909 | 13.894 |
| Observations | 11 032 884 | 11 032 884 |
| Number of individuals | 2 243 040 | 2 243 040 |

*Notes*: This table reports estimates of the regression coefficient on $treat_i \times post_{it}$ from equation (1) (column 1), and from an event-study specification that replaces $treat_i \times post_{it}$ with interactions for event-year 0 and event-year 1 (column 2). The sample is the same as in Table 1. Individuals receive separate identifiers when they appear in both the treated and control groups. Partner income is calculated as family income minus the individual's personal income. Outcome for $treat_i \times post_{it} = 0$ is provided for the absolute outcome in 10,000 SEK. Standard errors, clustered at the individual (experimental) level, are reported in parentheses.

*** $p < 0.01$, ** $p < 0.05$, * $p < 0.10$

Table 4. Mediating Impact of Pharmaceutical Treatments on IHS *Incomes of Family Members*: DDD Specification

| | Personal income | Partner income |
|---|---|---|
| | (1) | (2) |
| $treat_i \times post_{it} \times M_{dt}$ | 0.00483*** | 0.0212*** |
| | (0.000) | (0.000) |
| by event year | | |
| $treat_i \times M_{dt} \times event\ year\ 0$ | 0.00483*** | 0.0194*** |
| | (0.000) | (0.000) |
| $treat_i \times M_{dt} \times event\ year\ 1$ | 0.00486*** | 0.0230*** |
| | (0.000) | (0.000) |
| Outcome for $treat_i \times post_{it} = 0$ | 18.909 | 13.894 |
| Observations | 11 032 884 | 11 032 884 |
| Number of individuals | 2 243 040 | 2 243 040 |

*Notes*: This table reports estimates of the regression coefficient on $treat_i \times post_{it} \times M_{dt}$ from equation (2) (column 1), and from an event-study specification that replaces $treat_i \times post_{it} \times M_{dt}$ with interactions for event-year 0 and event-year 1 (column 2). The sample is the same as in Table 1. $M_{dt}$ denote pharmaceutical treatments that are a one-year lag of cumulative count of new molecular entities available in year *t* to treat disease *d* (see Section II.B for details). For treated individuals, $M_{dt}$ are fixed at the level of disease group and year of hospitalization. Individuals receive separate identifiers when they appear in both the treated and control groups. Partner income is calculated as family income minus the individual's personal income. Outcome for $treat_i \times post_{it} = 0$ is provided for the absolute outcome in 10,000 SEK. Standard errors, clustered at the individual (experimental) level, are reported in parentheses.

*** $p < 0.01$, ** $p < 0.05$, * $p < 0.10$

Table 5. Impact of a Health Shock and Mediating Impact of Pharmaceutical Treatments on *Outpatient Hospital Visits* and the Sources of *IHS Personal Incomes*: DD and DDD Specifications

| | Outpatient hospital visits | Wage | Sickness benefits | Unempl. benefits | Disability benefits | Early retirement benefits | Capital income |
|---|---|---|---|---|---|---|---|
| | (1) | (2) | (3) | (4) | (5) | (6) | (7) |
| (A) DD: | | | | | | | |
| $treat_i \times post_{it}$ | 1.647*** | -0.382*** | 2.497*** | 0.246*** | 0.176*** | 0.0603*** | 0.038*** |
| | (0.008) | (0.004) | (0.006) | (0.002) | (0.003) | (0.011) | (0.007) |
| (B) DDD: | | | | | | | |
| $treat_i \times post_{it} \times M_{dt}$ | -0.0293*** | 0.0166*** | -0.00221*** | -0.00471*** | -0.00295*** | -0.00357** | -0.000819 |
| | (0.000) | (0.000) | (0.000) | (0.000) | (0.000) | (0.001) | (0.000) |
| Outcome for $treat_i \times post_{it} = 0$ | 0.602 | 21.751 | 0.825 | 0.036 | 1.041 | 18.568 | 1.487 |
| Observations | 2 227 014 | 11 032 884 | 10 665 937 | 11 032 884 | 10 665 937 | 3 711 134 | 11 032 884 |
| Number of individuals | 625 410 | 2 243 040 | 2 242 971 | 2 243 040 | 2 242 971 | 975 643 | 2 243 040 |

*Notes*: This table reports estimates of the regression coefficient on $treat_i \times post_{it}$ from equation (1) (panel A) and on $treat_i \times post_{it} \times M_{dt}$ from equation (2) (panel B). The sample in column 1 comprises observations from 2001, the year the outpatient register became available. The sample for the columns 2, 4 and 7 is the same as in Table 1. The sample for columns 3 and 5 includes observations from the year of 1981 (when sickness and disability records became available). The sample for the column 5 includes records for the individuals between age 55 and 67 (eligible to benefits). $M_{dt}$ denote pharmaceutical treatments that are a one-year lag of cumulative count of new molecular entities available in year $t$ to treat disease $d$ (see Section II.B for details). For treated individuals, $M_{dt}$ are fixed at the level of disease group and year of hospitalization. Individuals receive separate identifiers when they appear in both the treated and control groups. Outcome for $treat_i \times post_{it} = 0$ is provided for the absolute outcome, in units for column 1 and in 10,000 SEK in columns 2−7. Standard errors, clustered at the individual (experimental) level, are reported in parentheses.

*** $p < 0.01$, ** $p < 0.05$, * $p < 0.10$

Table 6. Impact of a Health Shock and Mediating Impact of Pharmaceutical Treatments on *Partner Outpatient Hospital Visits* and the Sources of *IHS Partner Incomes*: DD and DDD Specifications

| | Outpatient hospital visits (mental diseases) | Income | Wage | Sickness benefits | Unemploy-ment benefits | Disability benefits | Early retirement benefits |
|---|---|---|---|---|---|---|---|
| | (1) | (2) | (3) | (4) | (5) | (6) | (7) |
| (A) DD: | | | | | | | |
| $treat_i \times post_{it}$ | 0.00407** | -0.226*** | -0.144*** | 0.0732*** | 0.00187 | 0.0237*** | 0.00978 |
| | (0.002) | (0.008) | (0.010) | (0.011) | (0.003) | (0.006) | (0.006) |
| (B) DDD: | | | | | | | |
| $treat_i \times post_{it} \times M_{dt}$ | -0.00249 | 0.0127*** | 0.00823*** | -0.00232* | -0.0000261 | 0.000515 | -0.00529 |
| | (0.005) | (0.000) | (0.001) | (0.001) | (0.000) | (0.000) | (0.004) |
| Outcome for $treat_i \times post_{it} = 0$ | 0.0292 | 12.608 | 10.303 | 1.557 | 0.113 | 1.171 | 4.322 |
| Observations | 1 721 416 | 2 119 318 | 2 119 318 | 2 119 318 | 2 119 318 | 2 119 318 | 789 847 |
| Number of individuals | 427 528 | 562 350 | 562 350 | 562 350 | 562 350 | 562 350 | 250 580 |

*Notes*: This table reports estimates of the regression coefficient on $treat_i \times post_{it}$ from equation (1) (panel A) and on $treat_i \times post_{it} \times M_{dt}$ from equation (2) (panel B). Column 1 includes observations from 2001 onward (when outpatient data became available) for mental disease visits (ICD-10 chapter F “Mental and Behavioral Disorders”). The sample for the columns 2-5 and 7 includes all partners observed from 1998 (when data on partner identifiers are available). The sample for the column 6 includes records for the individuals between age 55 and 67 (eligible to benefits). $M_{dt}$ denote pharmaceutical treatments that are a one-year lag of cumulative count of new molecular entities available in year $t$ to treat disease $d$ (see Section II.B for details). Income is a real disposable income recorded for the individual. For treated individuals, $M_{dt}$ are fixed at the level of disease group and year of hospitalization. Individuals receive separate identifiers when they appear in both the treated and control groups. Partner income is calculated as family income minus the individual’s personal income. Outcome for $treat_i \times post_{it} = 0$ is provided for the absolute outcome in 10,000 SEK. Standard errors, clustered at the individual (experimental) level, are reported in parentheses.

Table 7. Heterogeneous Impacts of a Health Shock and Mediating Impact of Pharmaceutical Treatments on IHS *Family, Personal, and Partner Income*: DD and DDD Specifications

| | Below age 60 | Above age 60, non-retired | Above age 60, retired | Cancer | Other than cancer |
|---|---|---|---|---|---|
| | (1) | (2) | (3) | (4) | (5) |
| | | | *I Family Income* | | |
| (A) DD: | | | | | |
| $treat_i \times post_{it}$ | -0.231*** | -0.823*** | -0.219*** | -0.928*** | -0.190*** |
| | (0.002) | (0.010) | (0.005) | (0.006) | (0.002) |
| (B) DDD: | | | | | |
| $treat_i \times post_{it} \times M_{dt}$ | 0.0122*** | 0.0420*** | 0.00346*** | 0.0496*** | 0.00452*** |
| | (0.000) | (0.000) | (0.000) | (0.000) | (0.000) |
| Outcome for $treat_i \times post_{it} = 0$ | 15.224 | 33.196 | 24.356 | 17.069 | 16.009 |
| | | | *II Personal Income* | | |
| (A) DD: | | | | | |
| $treat_i \times post_{it}$ | -0.0378*** | -0.0828*** | -0.0117*** | -0.151*** | -0.0302*** |
| | (0.002) | (0.007) | (0.002) | (0.004) | (0.002) |
| (B) DDD: | | | | | |
| $treat_i \times post_{it} \times M_{dt}$ | 0.00383*** | 0.0485*** | -0.000541 | 0.0287*** | 0.00205*** |
| | (0.000) | (0.000) | (0.000) | (0.001) | (0.000) |
| Outcome for $treat_i \times post_{it} = 0$ | 12.549 | 20.435 | 13.574 | 12.485 | 16.009 |
| | | | *III Partner Income* | | |
| (A) DD: | | | | | |
| $treat_i \times post_{it}$ | -0.354*** | -1.182*** | -0.318*** | -1.311*** | -0.295*** |
| | (0.005) | (0.018) | (0.012) | (0.012) | (0.004) |
| (B) DDD: | | | | | |
| $treat_i \times post_{it} \times M_{dt}$ | 0.0150*** | 0.621*** | 0.0413*** | 0.0587*** | 0.00528*** |
| | (0.000) | (0.002) | (0.001) | (0.002) | (0.000) |
| Outcome for $treat_i \times post_{it} = 0$ | 9.230 | 12.760 | 10.781 | 9.510 | 9.137 |
| Observations | 8 657 566 | 1 150 100 | 850 572 | 1 862 608 | 9 170 276 |
| Number of individuals | 1 912 041 | 420 724 | 287 225 | 378 866 | 1 864 174 |

*Notes*: This table reports estimates of the regression coefficient on $treat_i \times post_{it}$ from equation (1) (panel A) and on $treat_i \times post_{it} \times M_{dt}$ from equation (2) (panel B). The full sample is the same as in Table 1 and is further divided by age (and retirement status) or severity of disease. $M_{dt}$ denote pharmaceutical treatments that are a one-year lag of cumulative count of new molecular entities available in year $t$ to treat disease $d$ (see Section II.B for details). Income is a real disposable income recorded for the individual. For treated individuals, $M_{dt}$ are fixed at the level of disease group and year of hospitalization. Individuals receive separate identifiers when they appear in both the treated and control groups. Outcome for $treat_i \times post_{it} = 0$ is provided for the absolute outcome in 10,000 SEK. Standard errors, clustered at the individual (experimental) level, are reported in parentheses.

*** $p < 0.01$, ** $p < 0.05$, * $p < 0.10$

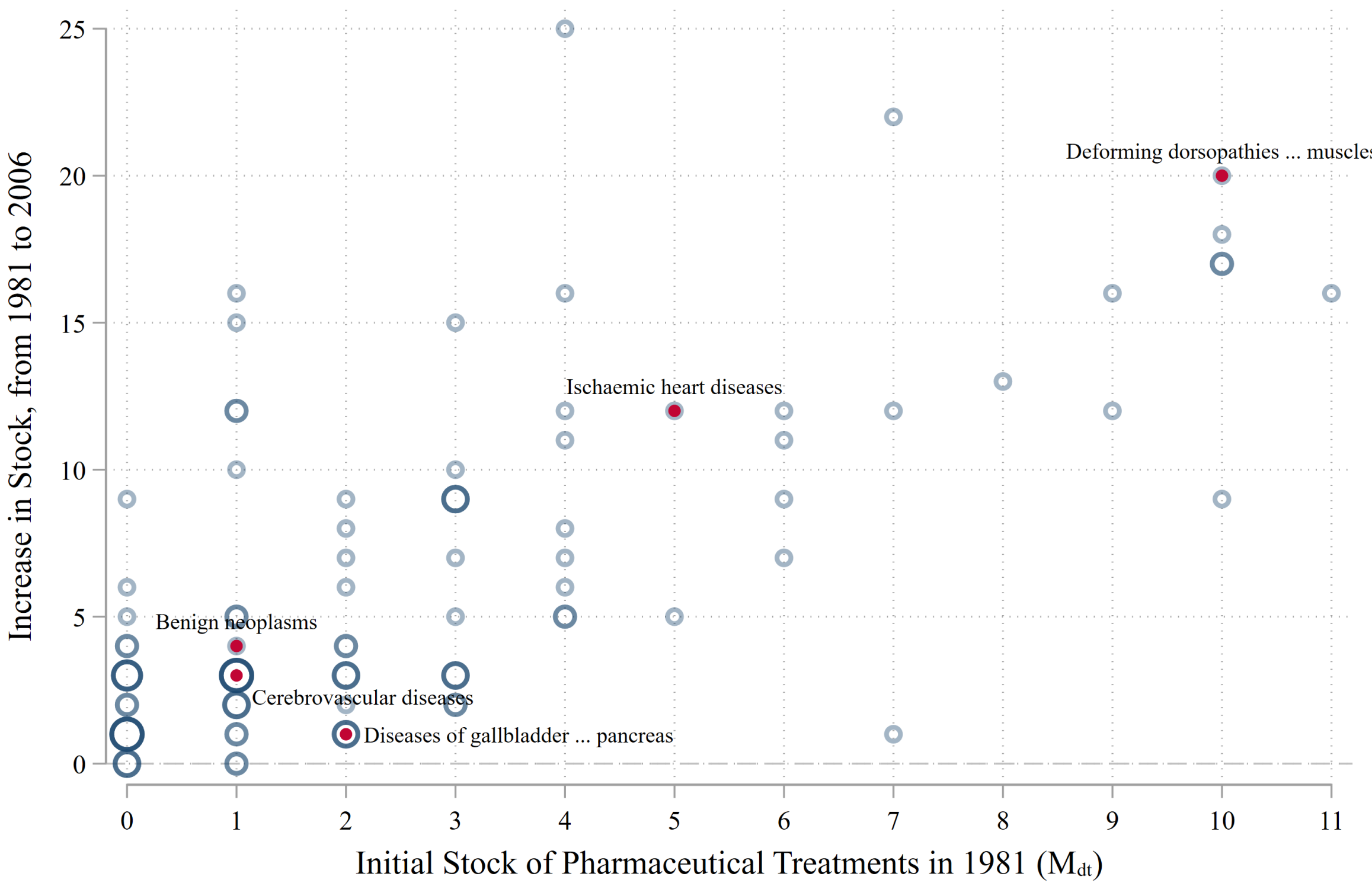

Figure 1. Distribution of Pharmaceutical Treatments Across the Sample Period.

*Notes*: This figure shows the range of pharmaceutical treatments in the sample by initial pharmaceutical treatment levels and their change between 1981 and 2006. Circle size corresponds to the number of disease groups in each cell (ranging from 1 to 5 out of 91 total groups). Red circles indicate the most significant disease groups by number of treated individuals. The main specification uses pharmaceutical treatment levels in the year an individual experiences a health shock (inpatient hospitalization) for a particular disease. $M_{dt}$ denote pharmaceutical preatments that are a one-year lag of cumulative count of new molecular entities available in year $t$ to treat disease $d$ (see Section II.B for details).

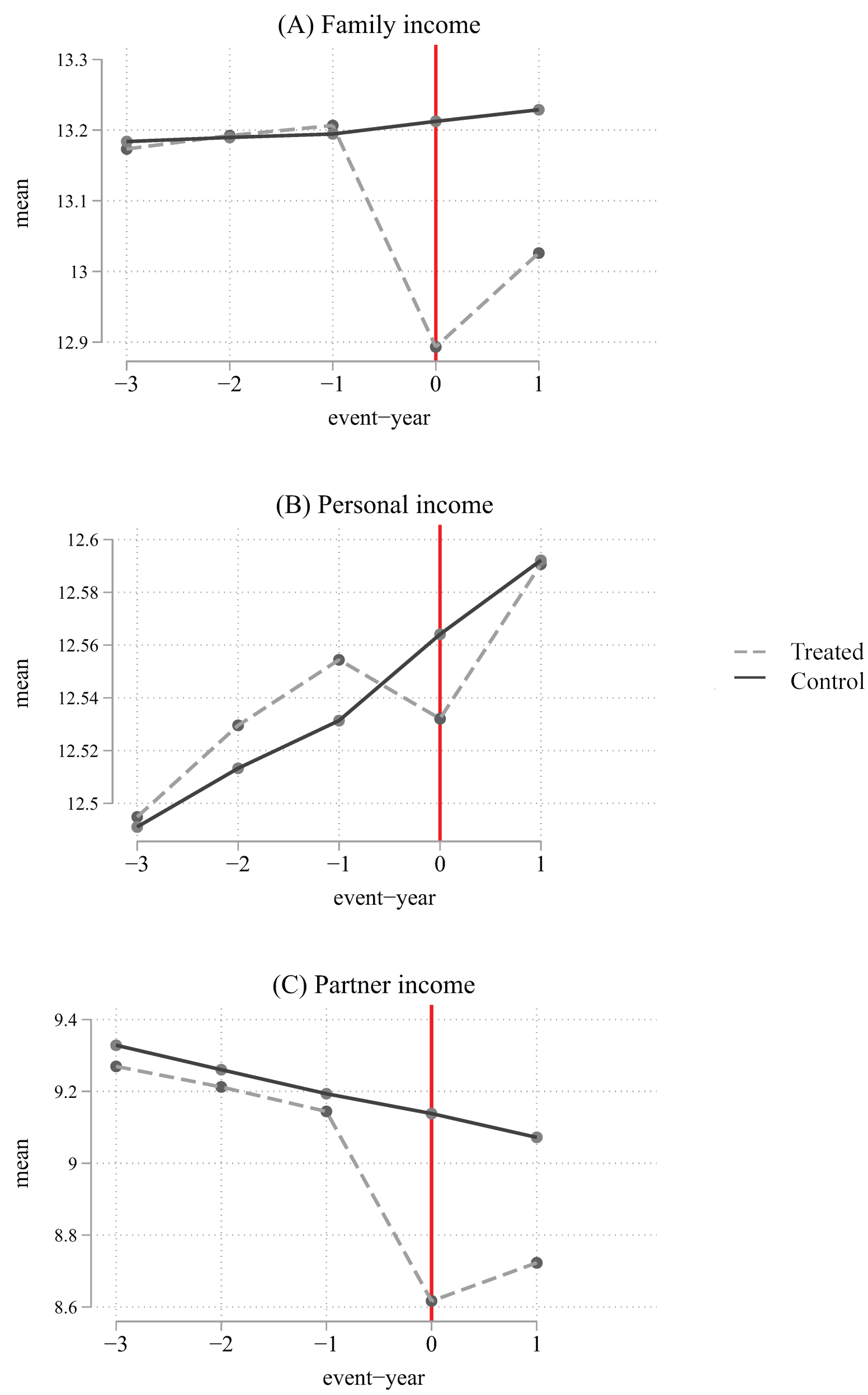

Figure 2. Development of Family, Personal, And Partner Income (IHS) by Event-Years for Treatment and Control Groups.

*Notes*: The figure shows sample means of the outcomes by treatment group and event year. The sample includes treated individuals—who experience a health shock (inpatient hospitalization)—and control individuals who experience the same health shock three years later. Each treated inidvidual is matched to the control individual within the same sex, disease group (91 groups), and year of hospitalization, as well as on the propensity score constructed from pre-event characteristics: year of birth, years of schooling, and earnings (IHS) (see Section III.B for details).

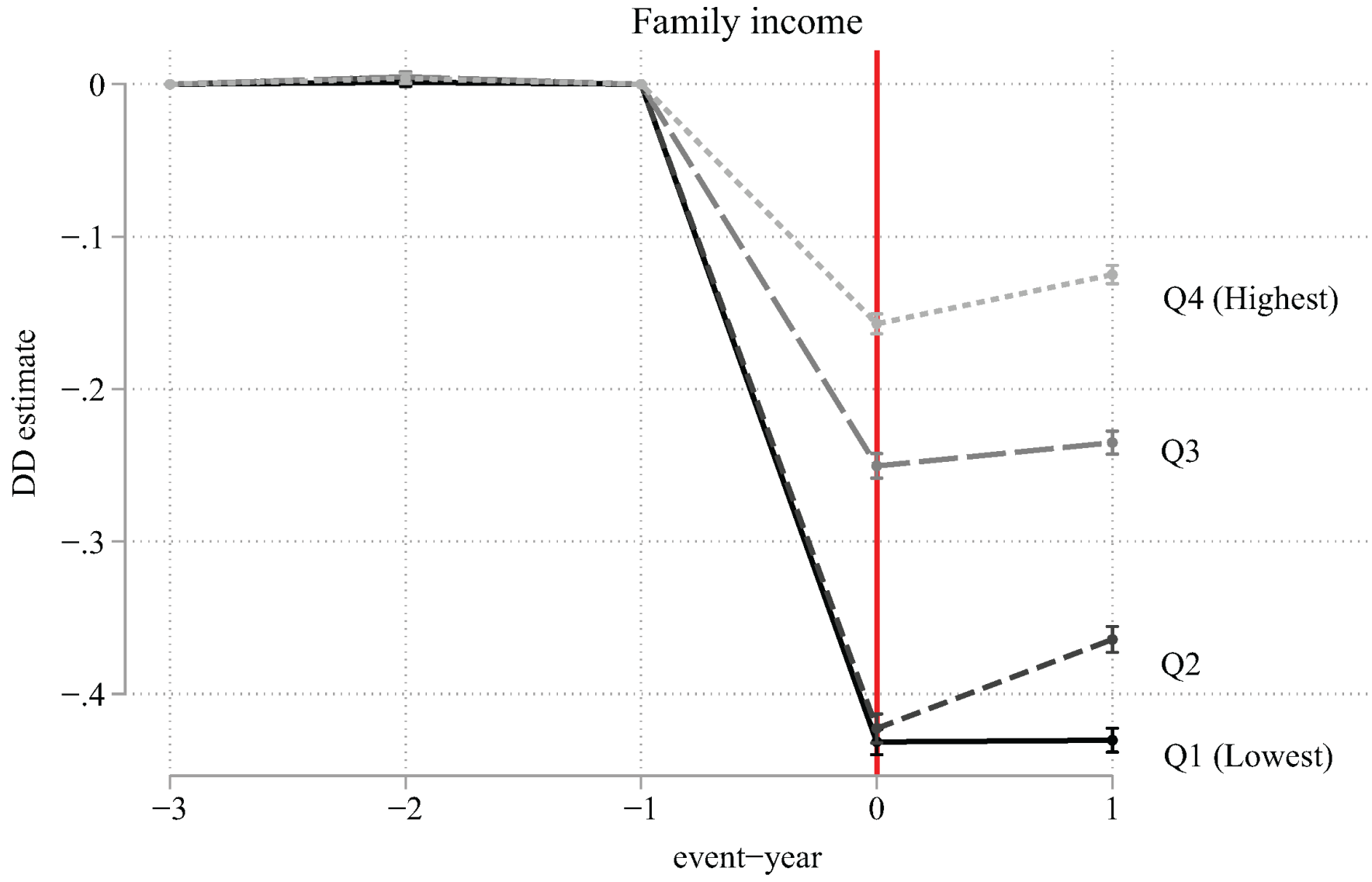

Figure 3. Impact of a Health Shock on IHS Family Income Across the Quartiles of Pharmaceutical Treatments: Baseline DD Specification

*Notes*: This figure reports estimates and 95% confidence intervals from the fully dynamic event-study specification of equation (1) across quartiles of medical innovation $M_{dt}$: Q1 (0–3 pharmaceutical treatments), Q2 (4–7 pharmaceutical treatments), Q3 (8–14 pharmaceutical treatments), and Q4 (15–30 pharmaceutical treatments). Event years -3 and -1 serve as reference categories to detect non-linearity in outcomes using t-tests and F-tests. $M_{dt}$ denote pharmaceutical preatments that are a one-year lag of cumulative count of new molecular entities available in year $t$ to treat disease $d$ (see Section II.B). For treated individuals, $M_{dt}$ is fixed at the disease group and year of hospitalization. Individuals who appear in both treated and control groups receive separate identifiers. Standard errors, clustered at the individual (experimental) level, are reported in parentheses.

*** $p < 0.01$, ** $p < 0.05$, * $p < 0.10$

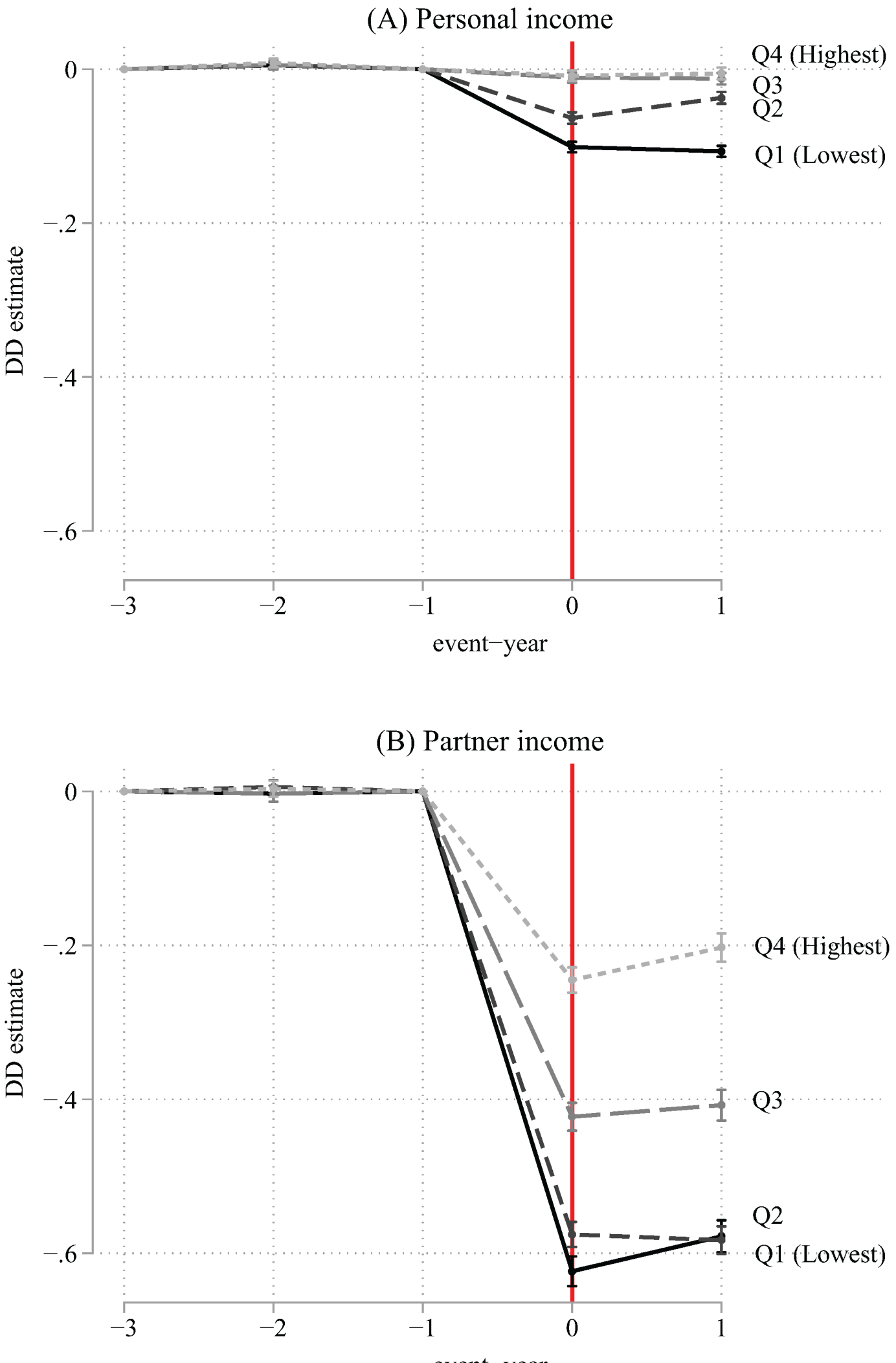

Figure 4. Impact of a Health Shock on IHS Personal and Partner Income Across the Quartiles of Pharmaceutical Treatments: Baseline DD Specification

This figure reports estimates and 95% confidence intervals from the fully dynamic event-study specification of equation (1) across quartiles of medical innovation $M_{dt}$: Q1 (0–3 pharmaceutical treatments), Q2 (4–7 pharmaceutical treatments), Q3 (8–14 pharmaceutical treatments), and Q4 (15–30 pharmaceutical treatments). Event years -3 and -1 serve as reference categories to detect non-linearity in outcomes using t-tests and F-tests. $M_{dt}$ denote pharmaceutical preatments that are a one-year lag of cumulative count of new molecular entities available in year $t$ to treat disease $d$ (see Section II.B). For treated individuals, $M_{dt}$ is fixed at the disease group and year of hospitalization. Individuals who appear in both treated and control groups receive separate identifiers. Standard errors, clustered at the individual (experimental) level, are reported in parentheses.

*** $p < 0.01$, ** $p < 0.05$, * $p < 0.10$

For Online Publication

Appendix to the study

# "HOUSEHOLD AND INDIVIDUAL ECONOMIC OUTCOMES OF DIFFERENT HEALTH SHOCKS: THE ROLE OF MEDICAL INNOVATIONS"

## Appendix A Disease groups used in the study

The ICD codes available in the inpatient register were aggregated into 91 disease groups, balancing clinically meaningful categories—as defined by Elixhauser, Steiner, and Palmer (2015)—with the availability and consistency of hospitalization coding over the study period. The final list was verified by health experts (Lindström and Rosvall 2019). Percentages represent prevalence in the sample. Disease groups are ordered in descending order by prevalence. The groups (ICD-10 Chapter, Disease Group, Prevalence in the Sample) are as follows:

| ICD Chapter Group | Disease Name | Prevalence |
|---|---|---|
| Circulatory Diseases | Ischaemic heart diseases | 8.1% |
| Cancer | Benign neoplasms | 5.7% |
| Musculoskeletal Diseases | Deforming dorsopathies, osteopathies and chondropathies. Disorders of muscles | 5.4% |
| Digestive Diseases | Diseases of gallbladder, biliary tract and pancreas | 5.2% |
| Circulatory Diseases | Cerebrovascular diseases | 4.1% |
| Circulatory Diseases | Diseases of veins, lymphatic vessels and lymph nodes, not elsewhere classified | 3.7% |
| Digestive Diseases | Hernia | 3.6% |
| Circulatory Diseases | Cardiac arrhythmias and heart failure | 3.2% |
| Digestive Diseases | Inflammatory bowel disease and other diseases of intestines | 3.0% |
| Cancer | Malignant neoplasm of breast | 2.8% |
| Digestive Diseases | Diseases of oesophagus, stomach and duodenum | 2.7% |
| Urinary Diseases | Diseases of female pelvic organs | 2.6% |
| Respiratory Diseases | Pneumonia, other acute lower respiratory infections and diseases of pleura | 2.5% |
| Musculoskeletal Diseases | Arthrosis and systemic connective tissue disorders | 2.5% |
| Urinary Diseases | Diseases of male genital organs | 2.1% |
| Respiratory Diseases | Diseases of upper respiratory tract | 2.1% |
| Musculoskeletal Diseases | Rheumatoid and juvenile arthritis. Gout | 2.0% |
| Mental Diseases | Mental and behavioural disorders due to use of alcohol and other substances | 2.0% |
| Cancer | In situ neoplasms | 1.8% |
| Digestive Diseases | Diseases of appendix | 1.8% |
| Urinary Diseases | Urolithiasis | 1.8% |
| Metabolic Diseases | Diabetes mellitus | 1.6% |
| Circulatory Diseases | Hypertensive diseases | 1.3% |
| Diseases of Sense Organs | Disorders of eyelid, lacrimal system and orbit, conjunctiva, sclera, cornea, iris, ciliary body, choroid and retina. | 1.3% |
| Urinary Diseases | Other diseases of the urinary system | 1.1% |
| Infectious and Parasitic Diseases | Bacterial diseases. Erysipelas. Meningitis | 1.1% |
| Metabolic Diseases | Disorders of thyroid gland | 1.1% |
| Cancer | Malignant neoplasm of small intestine, colon, rectosigmoid junction, rectum, anus and anal canal | 1.1% |
| Mental Diseases | Mood (affective) disorders | 1.1% |
| Diseases of Sense Organs | Diseases of inner ear | 1.0% |
| Mental Diseases | Neurotic, stress-related and somatoform disorders | 1.0% |
| Cancer | Malignant neoplasms of penis, prostate, testis and other male genital organs | 1.0% |
| Urinary Diseases | Glomerular diseases and renal tubulo-interstitial diseases. Renal failure | 0.9% |
| Circulatory Diseases | Diseases of arteries, arterioles and capillaries | 0.9% |

| Nervous Diseases | Migraine and other headache syndromes | 0.9% |
|---|---|---|
| Diseases of Sense Organs | Diseases of external and middle ear | 0.8% |
| Cancer | Malignant neoplasms of vulva, vagina, cervix uteri, corpus uteri and parts of uterus | 0.8% |
| Infectious and Parasitic Diseases | Intestinal infectious diseases | 0.8% |
| Nervous Diseases | Nerve, nerve root and plexus disorders, polyneuropathies and myneuropathies | 0.7% |
| Mental Diseases | Schizophrenia, schizotypal and delusional disorders | 0.7% |
| Diseases of Skin | Bullous disorders, dermatitis and eczema, urticaria and erythema | 0.6% |
| Circulatory Diseases | Endocarditis and myocarditis and cardiomyopathy | 0.6% |
| Respiratory Diseases | Asthma | 0.6% |
| Cancer | Malignant neoplasm of respiratory and intrathoracic organs | 0.6% |
| Circulatory Diseases | Pulmonary heart disease and diseases of pulmonary circulation | 0.5% |
| Respiratory Diseases | Chronic obstructive pulmonary disease and chronic bronchitis | 0.5% |
| Metabolic Diseases | Obesity and other hyperalimentation, metabolic disorders | 0.5% |
| Diseases of Skin | Infections of the skin | 0.4% |
| Cancer | Melanoma and other malignant neoplasms of skin | 0.4% |
| Cancer | Malignant neoplasm of bladder | 0.4% |
| Digestive Diseases | Diseases of liver | 0.4% |
| Mental Diseases | Organic, including symptomatic, mental disorders and Alzheimer disease. Systemic atrophies. | 0.4% |
| Cancer | Malignant neoplasms of ovary and placenta | 0.4% |
| Nervous Diseases | Sleep disorders | 0.4% |
| Infectious and Parasitic Diseases | Viral infections | 0.3% |
| Nervous Diseases | Epilepsy | 0.3% |
| Diseases of Bloodforming Organs | Nutritional anaemias | 0.3% |
| Diseases of Sense Organs | Disorders of globe, optical nerve and visual pathways, ocular muscles, accommodation and refraction, and blindness | 0.3% |
| Diseases of Sense Organs | Glaucoma | 0.3% |
| Infectious and Parasitic Diseases | Sexually transmitted diseases | 0.3% |
| Cancer | Malignant neoplasm of kidney, renal pelvis and ureter | 0.2% |
| Diseases of Sense Organs | Cataract, disorders of lens | 0.2% |
| Cancer | Non-Hodgkin's lymphoma | 0.2% |
| Musculoskeletal Diseases | Infectious arthropathies | 0.2% |
| Cancer | Malignant neoplasms of lip, oral cavity and pharynx | 0.2% |
| Cancer | Malignant neoplasm of stomach | 0.2% |
| Circulatory Diseases | Pericarditis | 0.2% |
| Cancer | Malignant neoplasms of eye and adnexa, meninges, brain, spinal cord, cranial nerves and other parts of central nervous system | 0.2% |
| Metabolic Diseases | Disorders of other endocrine glands | 0.2% |
| Cancer | Malignant neoplasm of pancreas | 0.2% |
| Cancer | Leukaemia | 0.1% |
| Diseases of Bloodforming Organs | Haemolytic anaemias | 0.1% |
| Nervous Diseases | Demyelinating diseases of the central nervous system | 0.1% |
| Nervous Diseases | Inflammatory diseases of the central nervous system | 0.1% |
| Diseases of Bloodforming Organs | Coagulation defects, purpura and other haemorrhagic conditions | 0.1% |
| Infectious and Parasitic Diseases | Viral hepatitis | 0.1% |
| Cancer | Malignant immunoproliferative diseases, multiple myeloma and malignant plasma cell neoplasms | 0.09% |
| Cancer | Malignant neoplasms of thyroid gland, adrenal gland, and other endocrine glands | 0.07% |
| Cancer | Malignant neoplasms of mesothelial and soft tissue | 0.07% |
| Cancer | Malignant neoplasm of oesophagus | 0.06% |
| Digestive Diseases | Diseases of peritoneum | 0.06% |
| Mental Diseases | Disorders of adult personality and behaviour | 0.05% |
| Infectious and Parasitic Diseases | Tuberculosis | 0.05% |
| Circulatory Diseases | Acute rheumatic fever and chronic rheumatic heart diseases | 0.05% |
| Cancer | Malignant neoplasm of liver and intrahepatic bile ducts | 0.05% |
| Cancer | Malignant neoplasm of gallbladder | 0.04% |
| Mental Diseases | Mental retardation. Disorders of psychological development, behavioral and emotional disorders | 0.03% |
| Infectious and Parasitic Diseases | Protozoal diseases | 0.02% |
| Cancer | Hodgkin’s disease | 0.02% |
| Cancer | Malignant neoplasm of bone and articular cartilage | 0.01% |
| Infectious and Parasitic Diseases | HIV | 0.002% |

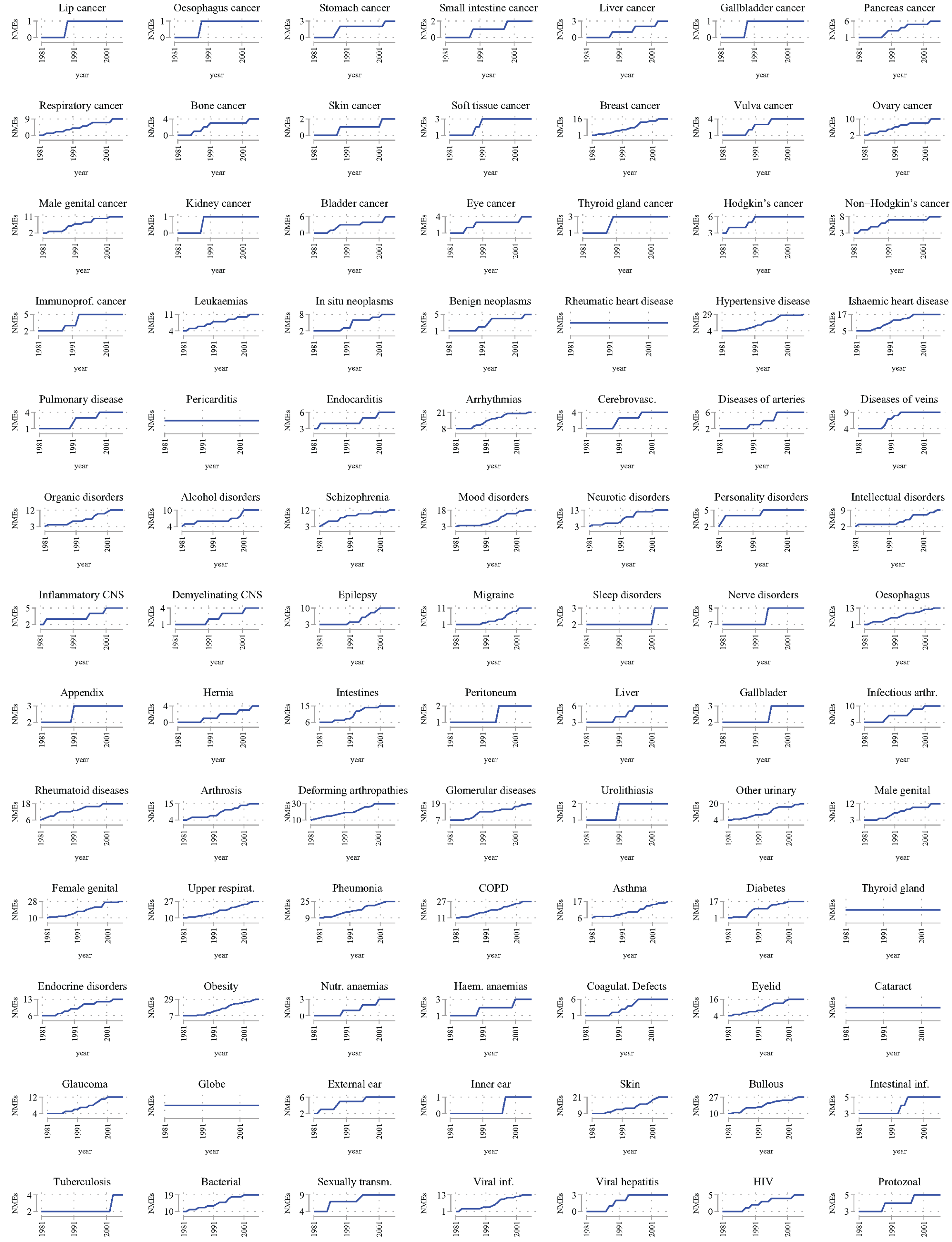

Figure A1 – Development of pharmaceutical preatments by disease over study period

*Notes*: Figure presents $M_{dt}$ by disease group over study period. $M_{dt}$ denote pharmaceutical preatments that are a one-year lag of cumulative count of new molecular entities—approvals net of disapprovals —available in year $t$ to treat disease $d$. See section II Data B Medical Innovations in the main body of the paper for more details.

## Appendix B Descriptive statistics and validation tests

Table B1 – Descriptive statistics for the estimation sample

| | **Observations** | **Mean** | **SD** |
|---|---|---|---|
| $M_{dt}$ | 11 032 884 | 8.960 | 7.458 |
| $treat_i \times post_{it}$ | 11 032 884 | 0.200 | 0.400 |
| family income (IHS) | 11 032 884 | 13.150 | 1.257 |
| personal income (IHS) | 11 032 884 | 12.539 | 1.639 |
| partner income (IHS) | 11 032 884 | 9.094 | 5.789 |
| personal wages (IHS) | 11 032 884 | 21.211 | 19.836 |
| personal sickness benefits (IHS) | 10 665 937 | 3.545 | 4.893 |
| personal unemployment benefits (IHS) | 11 032 884 | 0.231 | 1.489 |
| personal disability benefits (IHS) | 10 665 937 | 1.193 | 3.625 |
| personal early retirement benefits (IHS) | 3 711 134 | 11.229 | 1.856 |
| personal capital income (IHS) | 11 032 884 | -0.210 | 8.300 |
| outpatient hospital visits, from 2001 | 2 227 014 | 0.559 | 0.824 |
| partner income, from 1998 (IHS) | 2 119 318 | 12.617 | 1.793 |
| partner outpatient hospital visits (due to mental diseases, from 2001 | 1 721 416 | 0.028 | 0.140 |
| partner wages, from 1998 (IHS) | 2 119 318 | 10.253 | 5.621 |
| partner sickness benefits, from 1998 (IHS) | 2 119 318 | 1.566 | 3.854 |
| partner unemployment benefits, from 1998 (IHS) | 2 119 318 | 0.113 | 1.066 |
| partner disability benefits, from 1998 (IHS) | 2 119 318 | 1.198 | 3.630 |
| partner early retirement benefits, from 1998 (IHS) | 789 847 | 4.235 | 6.538 |

*Notes*: $M_{dt}$ denote pharmaceutical preatments that are a one-year lag of cumulative count of new molecular entities—approvals net of disapprovals —available in year $t$ to treat disease $d$. All absolute economic outcomes were adjusted for inflation, using 2021 as the base year. Records on sickness and disability benefits were available starting in 1981. Records on early retirement benefits were obtained for individuals aged 55–67 (eligible for such benefits). Records on partner sources of income were available from 1998 onward, linked through unique spouse identifiers.

Table B2 – Results of the *F*-test on non-linear pre-trends for family income (IHS) by disease group

| | Malignant neoplasms of lip, oral cavity and pharynx | Malignant neoplasm of oesophagus | Malignant neoplasm of stomach | Malignant neoplasm of small intestine, colon, rectosigmoid junction, rectum, anus and anal canal | Malignant neoplasm of liver and intrahepatic bile ducts | Malignant neoplasm of gallbladder | Malignant neoplasm of pancreas | Malignant neoplasm of respiratory and intrathoracic organs | Malignant neoplasm of bone and articular cartilage | Melanoma and other malignant neoplasms of skin | Malignant neoplasms of mesothelial and soft tissue | Malignant neoplasm of breast | Malignant neoplasms of vulva, vagina, cervix uteri, corpus uteri and parts of uterus |
|---|---|---|---|---|---|---|---|---|---|---|---|---|---|
| event year -2 | 0.027 | 0.001 | -0.006 | 0.007 | -0.014 | 0.016 | -0.001 | -0.002 | 0.023 | -0.001 | -0.025* | -0.004 | -0.007 |
| | (0.018) | (0.017) | (0.014) | (0.005) | (0.024) | (0.013) | (0.013) | (0.007) | (0.029) | (0.010) | (0.015) | (0.003) | (0.006) |
| event year 0 | 0.075*** | 0.015 | 0.032* | 0.013** | 0.036* | -0.013 | 0.006 | 0.029*** | 0.089* | 0.032*** | 0.023** | 0.025*** | 0.024*** |
| | (0.019) | (0.027) | (0.016) | (0.006) | (0.022) | (0.029) | (0.015) | (0.008) | (0.052) | (0.010) | (0.011) | (0.003) | (0.006) |
| event year 1 | 0.074*** | 0.027 | 0.040*** | 0.034*** | 0.057** | -0.004 | 0.007 | 0.049*** | 0.066 | 0.055*** | 0.042*** | 0.045*** | 0.032*** |
| | (0.024) | (0.024) | (0.015) | (0.007) | (0.029) | (0.018) | (0.018) | (0.009) | (0.066) | (0.012) | (0.012) | (0.004) | (0.007) |
| $DD_{idst}$ X event year -2 | 0.008 | -0.027 | 0.007 | -0.009 | 0.035 | -0.023 | -0.001 | 0.005 | -0.012 | 0.004 | 0.042* | -0.000 | 0.007 |
| | (0.024) | (0.030) | (0.018) | (0.007) | (0.031) | (0.018) | (0.018) | (0.011) | (0.034) | (0.012) | (0.024) | (0.004) | (0.008) |
| $DD_{idst}$ X event year 0 | -0.871*** | -3.746*** | -4.230*** | -1.400*** | -8.636*** | -6.972*** | -7.162*** | -4.763*** | -1.855*** | -0.653*** | -1.673*** | -0.210*** | -0.370*** |
| | (0.075) | (0.222) | (0.133) | (0.037) | (0.259) | (0.281) | (0.151) | (0.077) | (0.372) | (0.044) | (0.160) | (0.010) | (0.024) |
| $DD_{idst}$ X event year 1 | -1.882*** | -7.648*** | -6.024*** | -2.185*** | -9.979*** | -9.404*** | -10.571*** | -6.973*** | -3.673*** | -0.978*** | -2.063*** | -0.414*** | -0.894*** |
| | (0.106) | (0.308) | (0.178) | (0.046) | (0.502) | (0.449) | (0.232) | (0.106) | (0.514) | (0.051) | (0.181) | (0.013) | (0.036) |
| Constant | 13.086*** | 13.152*** | 13.186*** | 13.233*** | 13.115*** | 13.077*** | 13.199*** | 13.133*** | 13.167*** | 13.217*** | 13.285*** | 13.272*** | 13.133*** |
| | (0.014) | (0.034) | (0.020) | (0.006) | (0.034) | (0.039) | (0.020) | (0.011) | (0.065) | (0.008) | (0.026) | (0.002) | (0.005) |
| Number of IDs | 20,838 | 6,971 | 20,886 | 121,977 | 5,392 | 5,009 | 18,133 | 65,642 | 1,489 | 46,336 | 7,595 | 314,974 | 87,367 |
| R-squared | 0.083 | 0.399 | 0.333 | 0.113 | 0.634 | 0.546 | 0.571 | 0.381 | 0.193 | 0.044 | 0.115 | 0.015 | 0.041 |
| Obs. | 4,249 | 1,448 | 4,387 | 24,891 | 1,171 | 1,076 | 3,900 | 13,813 | 306 | 9,470 | 1,562 | 63,588 | 17,691 |
| *F-test: $DD_{idst}$ X event year -2 =0* | *0.741* | *0.367* | *0.689* | *0.222* | *0.264* | *0.220* | *0.961* | *0.622* | *0.721* | *0.765* | *0.0753* | *0.933* | *0.363* |
| *Standardised difference* | *0.014* | *0.025* | *0.011* | *0.013* | *0.077* | *0.031* | *0.014* | *0.013* | *0.017* | *0.035* | *0.037* | *0.001* | *0.005* |

Table B2 cont'd

| | Malignant neoplasms of ovary and placenta | Malignant neoplasms of penis, prostate, testis and other male genital organs | Malignant neoplasm of kidney, renal pelvis and ureter | Malignant neoplasm of bladder | Malignant neoplasms of eye and adnexa, meninges, brain, spinal cord, cranial nerves and other parts of central nervous system | Malignant neoplasms of thyroid gland, adrenal gland, and other endocrine glands | Hodgkin's disease | Non-Hodgkin's lymphoma | Malignant immunoproliferative diseases, multiple myeloma and malignant plasma cell neoplasms | Leukaemia | In situ neoplasms | Benign neoplasms | Acute rheumatic fever and chronic rheumatic heart diseases |
|---|---|---|---|---|---|---|---|---|---|---|---|---|---|
| event year -2 | 0.001 | -0.002 | -0.008 | 0.001 | -0.007 | 0.009 | -0.024 | -0.016* | -0.032 | 0.003 | 0.000 | -0.004** | 0.010 |
| | (0.005) | (0.004) | (0.011) | (0.008) | (0.012) | (0.008) | (0.039) | (0.009) | (0.022) | (0.017) | (0.004) | (0.002) | (0.012) |
| event year 0 | 0.020** | 0.025*** | 0.030*** | 0.049*** | 0.030* | 0.029*** | 0.040 | 0.028** | 0.029** | 0.044*** | 0.039*** | 0.036*** | -0.025 |
| | (0.008) | (0.006) | (0.011) | (0.010) | (0.017) | (0.010) | (0.038) | (0.012) | (0.014) | (0.015) | (0.004) | (0.002) | (0.037) |
| event year 1 | 0.033*** | 0.051*** | 0.051*** | 0.047*** | 0.037** | 0.038*** | -0.038 | 0.030* | 0.019 | 0.046*** | 0.051*** | 0.054*** | 0.031 |
| | (0.010) | (0.006) | (0.011) | (0.012) | (0.017) | (0.013) | (0.043) | (0.015) | (0.022) | (0.015) | (0.005) | (0.002) | (0.031) |
| $DD_{idst}$ X event year -2 | -0.006 | -0.003 | 0.011 | 0.010 | 0.020 | -0.002 | 0.102 | 0.025* | 0.036 | 0.011 | 0.010* | 0.008*** | -0.024 |
| | (0.009) | (0.007) | (0.015) | (0.012) | (0.018) | (0.011) | (0.067) | (0.014) | (0.025) | (0.022) | (0.005) | (0.003) | (0.037) |
| $DD_{idst}$ X event year 0 | -0.882*** | -0.410*** | -1.935*** | -0.525*** | -2.780*** | -0.549*** | -0.329* | -1.130*** | -1.148*** | -1.974*** | -1.004*** | -0.025*** | -0.241** |
| | (0.053) | (0.023) | (0.090) | (0.039) | (0.120) | (0.096) | (0.186) | (0.075) | (0.119) | (0.117) | (0.025) | (0.004) | (0.094) |
| $DD_{idst}$ X event year 1 | -2.120*** | -0.817*** | -2.614*** | -0.932*** | -5.061*** | -0.450*** | -0.784*** | -1.852*** | -1.902*** | -3.358*** | -1.294*** | -0.027*** | -0.245*** |
| | (0.079) | (0.030) | (0.105) | (0.051) | (0.165) | (0.082) | (0.244) | (0.094) | (0.152) | (0.153) | (0.028) | (0.004) | (0.081) |
| Constant | 13.187*** | 13.423*** | 13.231*** | 13.207*** | 13.260*** | 13.260*** | 13.130*** | 13.244*** | 13.236*** | 13.258*** | 13.220*** | 13.292*** | 13.021*** |
| | (0.010) | (0.004) | (0.014) | (0.007) | (0.019) | (0.015) | (0.034) | (0.013) | (0.020) | (0.020) | (0.004) | (0.001) | (0.015) |
| Number of IDs | 39,213 | 110,851 | 27,074 | 46,275 | 20,133 | 7,814 | 1,653 | 24,594 | 10,070 | 16,168 | 203,555 | 632,599 | 5,169 |
| R-squared | 0.110 | 0.035 | 0.143 | 0.040 | 0.262 | 0.028 | 0.039 | 0.095 | 0.096 | 0.174 | 0.066 | 0.002 | 0.010 |
| Obs. | 7,983 | 22,369 | 5,557 | 9,384 | 4,196 | 1,597 | 339 | 5,024 | 2,056 | 3,328 | 41,562 | 127,919 | 1,065 |
| *F-test: $DD_{idst}$ X event year -2 =0* | *0.483* | *0.605* | *0.460* | *0.365* | *0.273* | *0.832* | *0.128* | *0.0747* | *0.158* | *0.629* | *0.0574* | *0.00299* | *0.512* |
| *Standardised difference* | *0.001* | *0.015* | *0.011* | *0.016* | *0.015* | *0.010* | *0.170* | *0.004* | *0.046* | *0.037* | *0.005* | *0.002* | *0.057* |

Table B2 cont'd

| | Hypertensive diseases | Ischaemic heart diseases | Pulmonary heart disease and diseases of pulmonary circulation | Pericarditis | Endocarditis and myocarditis and cardiomyopathy | Cardiac arrhythmias and heart failure | Cerebrovascular diseases | Diseases of arteries, arterioles and capillaries | Diseases of veins, lymphatic vessels and lymph nodes, not elsewhere classified | Organic, including symptomatic, mental disorders and Alzheimer disease. Systemic atrophies. | Mental and behavioural disorders due to use of alcohol and other substances | Schizophrenia, schizotypal and delusional disorders | Mood (affective) disorders |
|---|---|---|---|---|---|---|---|---|---|---|---|---|---|
| event year -2 | -0.002 | -0.001 | 0.007 | -0.007 | -0.005 | 0.000 | -0.002 | -0.005 | 0.000 | -0.007 | -0.005 | -0.006 | 0.000 |
| | (0.006) | (0.002) | (0.007) | (0.010) | (0.009) | (0.003) | (0.003) | (0.007) | (0.003) | (0.009) | (0.007) | (0.009) | (0.006) |
| event year 0 | 0.025*** | 0.030*** | 0.040*** | -0.008 | 0.032*** | 0.035*** | 0.025*** | 0.019*** | 0.030*** | 0.000 | -0.006 | -0.011 | 0.019*** |
| | (0.006) | (0.002) | (0.009) | (0.014) | (0.009) | (0.004) | (0.003) | (0.007) | (0.003) | (0.010) | (0.008) | (0.011) | (0.007) |
| event year 1 | 0.047*** | 0.043*** | 0.060*** | -0.008 | 0.057*** | 0.054*** | 0.043*** | 0.037*** | 0.050*** | -0.008 | -0.012 | 0.002 | 0.004 |
| | (0.007) | (0.002) | (0.009) | (0.017) | (0.009) | (0.004) | (0.004) | (0.008) | (0.003) | (0.014) | (0.009) | (0.013) | (0.009) |
| $DD_{idst}$ X event year -2 | -0.005 | 0.006** | -0.004 | -0.007 | -0.008 | 0.003 | 0.005 | 0.007 | 0.002 | 0.010 | 0.008 | -0.002 | 0.007 |
| | (0.008) | (0.003) | (0.011) | (0.020) | (0.012) | (0.004) | (0.004) | (0.009) | (0.004) | (0.013) | (0.010) | (0.013) | (0.009) |
| $DD_{idst}$ X event year 0 | -0.091*** | -0.576*** | -0.673*** | -0.264*** | -0.424*** | -0.320*** | -1.057*** | -0.830*** | -0.096*** | -0.729*** | -0.159*** | -0.113*** | -0.222*** |
| | (0.012) | (0.009) | (0.039) | (0.050) | (0.030) | (0.012) | (0.017) | (0.033) | (0.007) | (0.050) | (0.015) | (0.023) | (0.018) |
| $DD_{idst}$ X event year 1 | -0.113*** | -0.263*** | -0.358*** | -0.167*** | -0.278*** | -0.234*** | -0.431*** | -0.407*** | -0.093*** | -0.854*** | -0.156*** | -0.039* | -0.123*** |
| | (0.013) | (0.006) | (0.027) | (0.043) | (0.024) | (0.010) | (0.010) | (0.021) | (0.006) | (0.054) | (0.016) | (0.022) | (0.015) |
| Constant | 13.148*** | 13.188*** | 13.235*** | 13.302*** | 13.219*** | 13.233*** | 13.180*** | 13.103*** | 13.163*** | 13.117*** | 12.769*** | 12.613*** | 13.121*** |
| | (0.002) | (0.001) | (0.006) | (0.008) | (0.005) | (0.002) | (0.002) | (0.005) | (0.001) | (0.008) | (0.003) | (0.004) | (0.003) |
| Number of IDs | 146,581 | 905,759 | 58,197 | 19,913 | 69,909 | 353,852 | 457,710 | 101,846 | 403,297 | 40,674 | 217,257 | 77,074 | 118,220 |
| R-squared | 0.001 | 0.025 | 0.030 | 0.008 | 0.014 | 0.011 | 0.056 | 0.040 | 0.002 | 0.043 | 0.003 | 0.001 | 0.006 |
| Obs. | 29,722 | 183,512 | 11,838 | 4,046 | 14,178 | 71,599 | 93,037 | 20,744 | 82,385 | 8,290 | 44,544 | 15,891 | 24,061 |
| *F-test: $DD_{idst}$ X event year -2 =0* | *0.530* | *0.0375* | *0.676* | *0.712* | *0.504* | *0.532* | *0.234* | *0.421* | *0.658* | *0.448* | *0.391* | *0.900* | *0.396* |
| *Standardised difference* | *0.020* | *0.008* | *0.015* | *0.022* | *0.017* | *0.005* | *0.004* | *0.027* | *0.008* | *0.019* | *0.070* | *0.040* | *0.020* |

Table B2
cont'd

| | Neurotic, stress-related and somatoform disorders | Disorders of adult personality and behaviour | Mental retardation. Disorders of psychological development, behavioral and emotional disorders | Inflammatory diseases of the central nervous system | Demyelinating diseases of the central nervous system | Epilepsy | Migraine and other headache syndromes | Sleep disorders | Nerve, nerve root and plexus disorders, polyneuropathies and myneuropathies | Diseases of oesophagus, stomach and duodenum | Diseases of appendix | Hernia | Inflammatory bowel disease and other diseases of intestines |
|---|---|---|---|---|---|---|---|---|---|---|---|---|---|
| event year -2 | -0.007 | 0.078** | 0.030 | 0.012 | 0.004 | -0.025* | 0.005 | -0.005 | 0.006 | -0.005 | -0.004 | -0.001 | 0.000 |
| | (0.007) | (0.038) | (0.039) | (0.019) | (0.008) | (0.013) | (0.006) | (0.011) | (0.005) | (0.004) | (0.004) | (0.003) | (0.003) |
| event year 0 | 0.011 | 0.016 | 0.041 | 0.059*** | 0.012 | 0.012 | 0.026*** | 0.085*** | 0.025*** | 0.024*** | 0.035*** | 0.034*** | 0.037*** |
| | (0.008) | (0.051) | (0.060) | (0.021) | (0.009) | (0.011) | (0.007) | (0.012) | (0.007) | (0.004) | (0.005) | (0.003) | (0.004) |
| event year 1 | 0.020** | 0.008 | 0.055 | 0.091*** | 0.005 | 0.015 | 0.049*** | 0.123*** | 0.051*** | 0.039*** | 0.071*** | 0.052*** | 0.059*** |
| | (0.009) | (0.056) | (0.060) | (0.021) | (0.017) | (0.013) | (0.007) | (0.013) | (0.007) | (0.005) | (0.005) | (0.003) | (0.004) |
| $DD_{idst}$ X event year -2 | 0.004 | -0.071 | -0.100 | 0.005 | 0.003 | 0.022 | -0.003 | 0.010 | -0.009 | 0.008 | -0.003 | 0.006 | 0.004 |
| | (0.010) | (0.061) | (0.070) | (0.024) | (0.012) | (0.016) | (0.008) | (0.015) | (0.008) | (0.005) | (0.006) | (0.004) | (0.005) |
| $DD_{idst}$ X event year 0 | -0.184*** | -0.253** | -0.244** | -0.509*** | -0.094*** | -0.237*** | -0.060*** | -0.054*** | -0.087*** | -0.117*** | -0.024*** | -0.026*** | -0.087*** |
| | (0.017) | (0.102) | (0.117) | (0.072) | (0.035) | (0.031) | (0.014) | (0.019) | (0.016) | (0.009) | (0.008) | (0.005) | (0.008) |
| $DD_{idst}$ X event year 1 | -0.139*** | -0.108 | -0.093 | -0.331*** | -0.046 | -0.309*** | -0.072*** | -0.062*** | -0.143*** | -0.140*** | -0.033*** | -0.034*** | -0.096*** |
| | (0.015) | (0.091) | (0.097) | (0.055) | (0.032) | (0.036) | (0.013) | (0.021) | (0.018) | (0.010) | (0.009) | (0.006) | (0.008) |
| Constant | 13.052*** | 12.580*** | 12.470*** | 13.232*** | 13.229*** | 13.018*** | 13.210*** | 13.366*** | 13.171*** | 13.115*** | 13.282*** | 13.160*** | 13.240*** |
| | (0.003) | (0.020) | (0.022) | (0.012) | (0.006) | (0.006) | (0.003) | (0.004) | (0.003) | (0.002) | (0.002) | (0.001) | (0.001) |
| Number of IDs | 108,296 | 5,852 | 3,279 | 14,324 | 15,157 | 35,867 | 95,877 | 38,866 | 81,078 | 294,241 | 199,061 | 394,345 | 334,767 |
| R-squared | 0.005 | 0.004 | 0.004 | 0.018 | 0.002 | 0.012 | 0.001 | 0.005 | 0.002 | 0.002 | 0.002 | 0.001 | 0.001 |
| Obs. | 22,265 | 1,226 | 672 | 2,916 | 3,102 | 7,378 | 19,540 | 7,854 | 16,578 | 59,941 | 40,485 | 80,213 | 67,903 |
| *F-test: $DD_{idst}$ X event year -2 =0* | *0.696* | *0.247* | *0.157* | *0.846* | *0.809* | *0.159* | *0.709* | *0.496* | *0.301* | *0.125* | *0.651* | *0.115* | *0.406* |
| *Standardised difference* | *0.009* | *0.111* | *0.102* | *0.020* | *0.036* | *0.060* | *0.010* | *0.020* | *0.008* | *0.001* | *0.008* | *0.010* | *0.001* |

Table B2
cont'd

| | Diseases of peritoneum | Diseases of liver | Diseases of gallbladder, biliary tract and pancreas | Infectious arthropathies | Rheumatoid and juvenile arthritis. Gout | Arthrosis and systemic connective tissue disorders | Deforming dorsopathies, osteopathies and chondropathies. Disorders of muscles | Glomerular diseases and renal tubulo-interstitial diseases. Renal failure | Urolithiasis | Other diseases of the urinary system | Diseases of male genital organs | Diseases of female pelvic organs | Diseases of upper respiratory tract |
|---|---|---|---|---|---|---|---|---|---|---|---|---|---|
| event year -2 | 0.010 | -0.001 | 0.002 | 0.005 | -0.000 | -0.002 | -0.003 | -0.002 | -0.001 | -0.004 | 0.006* | 0.002 | -0.003 |
| | (0.028) | (0.010) | (0.002) | (0.010) | (0.003) | (0.003) | (0.002) | (0.006) | (0.004) | (0.006) | (0.003) | (0.003) | (0.004) |
| event year 0 | 0.038** | -0.011 | 0.033*** | 0.040*** | 0.017*** | 0.038*** | 0.037*** | 0.035*** | 0.040*** | 0.045*** | 0.024*** | 0.044*** | 0.043*** |
| | (0.017) | (0.013) | (0.003) | (0.012) | (0.003) | (0.004) | (0.003) | (0.007) | (0.005) | (0.005) | (0.004) | (0.004) | (0.004) |
| event year 1 | 0.072*** | 0.003 | 0.054*** | 0.056*** | 0.020*** | 0.065*** | 0.055*** | 0.047*** | 0.060*** | 0.063*** | 0.038*** | 0.080*** | 0.063*** |
| | (0.027) | (0.014) | (0.003) | (0.013) | (0.003) | (0.004) | (0.003) | (0.008) | (0.005) | (0.006) | (0.004) | (0.003) | (0.004) |
| $DD_{idst}$ X event year -2 | 0.008 | -0.009 | 0.000 | -0.005 | 0.002 | 0.004 | 0.006* | 0.013 | 0.004 | 0.004 | -0.005 | -0.003 | 0.005 |
| | (0.033) | (0.015) | (0.003) | (0.015) | (0.004) | (0.004) | (0.003) | (0.008) | (0.006) | (0.007) | (0.004) | (0.004) | (0.005) |
| $DD_{idst}$ X event year 0 | -0.494*** | -1.750*** | -0.091*** | -0.034 | -0.038*** | -0.052*** | -0.042*** | -0.174*** | -0.024*** | -0.054*** | -0.040*** | -0.017*** | -0.040*** |
| | (0.107) | (0.067) | (0.006) | (0.024) | (0.006) | (0.007) | (0.005) | (0.018) | (0.008) | (0.010) | (0.007) | (0.006) | (0.007) |
| $DD_{idst}$ X event year 1 | -0.328*** | -1.083*** | -0.115*** | -0.028 | -0.052*** | -0.065*** | -0.049*** | -0.181*** | -0.042*** | -0.086*** | -0.059*** | -0.020*** | -0.044*** |
| | (0.079) | (0.052) | (0.006) | (0.024) | (0.007) | (0.007) | (0.005) | (0.018) | (0.009) | (0.012) | (0.008) | (0.006) | (0.007) |
| Constant | 13.240*** | 13.089*** | 13.229*** | 13.272*** | 13.136*** | 13.281*** | 13.250*** | 13.190*** | 13.227*** | 13.300*** | 13.208*** | 13.380*** | 13.264*** |
| | (0.016) | (0.010) | (0.001) | (0.005) | (0.001) | (0.001) | (0.001) | (0.003) | (0.002) | (0.002) | (0.001) | (0.001) | (0.001) |
| Number of IDs | 6,095 | 45,333 | 575,124 | 24,219 | 219,899 | 277,086 | 596,224 | 102,261 | 197,620 | 120,582 | 230,884 | 288,248 | 225,181 |
| R-squared | 0.020 | 0.097 | 0.002 | 0.001 | 0.001 | 0.001 | 0.001 | 0.004 | 0.001 | 0.001 | 0.001 | 0.004 | 0.001 |
| Obs. | 1,250 | 9,346 | 116,465 | 4,911 | 44,426 | 55,872 | 121,225 | 20,832 | 40,276 | 24,398 | 46,599 | 58,333 | 45,895 |
| *F-test: $DD_{idst}$ X event year -2 =0* | *0.807* | *0.555* | *0.897* | *0.739* | *0.618* | *0.377* | *0.0729* | *0.108* | *0.474* | *0.558* | *0.249* | *0.408* | *0.320* |
| *Standardised difference* | *0.105* | *0.045* | *0.003* | *0.002* | *0.011* | *0.008* | *0.004* | *0.009* | *0.015* | *0.011* | *0.010* | *0.007* | *0.015* |

Table B2 cont'd

| | Pneumonia, other acute lower respiratory infections and diseases of pleura | Chronic obstructive pulmonary disease and chronic bronchitis | Asthma | Diabetes mellitus | Disorders of thyroid gland | Disorders of other endocrine glands | Obesity and other hyperalimentation, metabolic disorders | Nutritional anaemias | Haemolytic anaemias | Coagulation defects, purpura and other haemorrhagic conditions | Disorders of eyelid, lacrimal system and orbit, conjunctiva, sclera, cornea, iris, ciliary body, choroid and retina. | Cataract, disorders of lens | Glaucoma |
|---|---|---|---|---|---|---|---|---|---|---|---|---|---|
| event year -2 | 0.003 | 0.000 | 0.010 | -0.003 | 0.002 | 0.002 | -0.011 | -0.006 | 0.013 | 0.001 | 0.002 | -0.004 | 0.011* |
| | (0.004) | (0.008) | (0.006) | (0.006) | (0.005) | (0.017) | (0.009) | (0.010) | (0.018) | (0.014) | (0.005) | (0.008) | (0.006) |
| event year 0 | 0.032*** | 0.033*** | 0.025*** | 0.046*** | 0.033*** | 0.043*** | 0.034*** | 0.011 | 0.021 | 0.025 | 0.029*** | 0.013 | 0.020** |
| | (0.004) | (0.009) | (0.008) | (0.006) | (0.005) | (0.012) | (0.009) | (0.013) | (0.018) | (0.018) | (0.006) | (0.008) | (0.009) |
| event year 1 | 0.052*** | 0.037*** | 0.043*** | 0.051*** | 0.049*** | 0.055*** | 0.049*** | 0.035** | 0.056*** | 0.029 | 0.060*** | 0.028** | 0.024*** |
| | (0.005) | (0.010) | (0.009) | (0.006) | (0.006) | (0.014) | (0.011) | (0.015) | (0.022) | (0.022) | (0.006) | (0.011) | (0.009) |
| $DD_{idst}$ X event year -2 | -0.003 | 0.012 | -0.007 | 0.018** | 0.000 | 0.006 | 0.009 | -0.005 | -0.026 | -0.005 | -0.007 | -0.001 | -0.002 |
| | (0.005) | (0.010) | (0.009) | (0.007) | (0.006) | (0.020) | (0.012) | (0.015) | (0.024) | (0.021) | (0.007) | (0.013) | (0.010) |
| $DD_{idst}$ X event year 0 | -0.396*** | -0.340*** | -0.148*** | -0.154*** | -0.044*** | -0.126*** | -0.149*** | -0.138*** | -0.640*** | -0.508*** | -0.015 | 0.003 | -0.037** |
| | (0.015) | (0.029) | (0.021) | (0.013) | (0.009) | (0.035) | (0.023) | (0.030) | (0.075) | (0.072) | (0.009) | (0.016) | (0.016) |
| $DD_{idst}$ X event year 1 | -0.395*** | -0.428*** | -0.160*** | -0.165*** | -0.041*** | -0.150*** | -0.150*** | -0.261*** | -0.712*** | -0.462*** | -0.037*** | -0.056** | -0.066*** |
| | (0.014) | (0.033) | (0.021) | (0.014) | (0.009) | (0.035) | (0.023) | (0.037) | (0.077) | (0.066) | (0.010) | (0.024) | (0.021) |
| Constant | 13.170*** | 12.990*** | 13.068*** | 13.024*** | 13.212*** | 13.192*** | 13.186*** | 13.051*** | 13.109*** | 13.199*** | 13.238*** | 12.994*** | 13.055*** |
| | (0.002) | (0.005) | (0.004) | (0.002) | (0.002) | (0.007) | (0.004) | (0.006) | (0.012) | (0.012) | (0.002) | (0.004) | (0.003) |
| Number of IDs | 276,503 | 56,026 | 63,493 | 177,291 | 119,112 | 19,171 | 49,797 | 35,400 | 15,735 | 13,292 | 140,473 | 25,593 | 28,879 |
| R-squared | 0.016 | 0.017 | 0.004 | 0.003 | 0.001 | 0.003 | 0.003 | 0.006 | 0.032 | 0.024 | 0.001 | 0.001 | 0.001 |
| Obs. | 56,286 | 11,407 | 12,991 | 36,209 | 24,285 | 3,912 | 10,179 | 7,218 | 3,223 | 2,721 | 28,444 | 5,192 | 5,861 |
| *F-test: $DD_{idst}$ X event year -2 =0* | *0.598* | *0.234* | *0.468* | *0.0120* | *0.940* | *0.751* | *0.462* | *0.715* | *0.270* | *0.807* | *0.339* | *0.960* | *0.839* |
| *Standardised difference* | *0.004* | *0.021* | *0.031* | *0.012* | *0.004* | *0.032* | *0.007* | *0.010* | *0.007* | *0.002* | *0.007* | *0.072* | *0.009* |

Table B2
cont'd

| | Disorders of globe, optical nerve and visual pathways, ocular muscles, accommodation and refraction, and blindness | Diseases of external and middle ear | Diseases of inner ear | Infections of the skin | Bullous disorders, dermatitis and eczema, urticaria and erythema | Intestinal infectious diseases | Tuberculosis | Bacterial diseases. Erysipelas. Meningitis | Sexually transmitted diseases | Viral infections | Viral hepatitis | HIV | Protozoal diseases |
|---|---|---|---|---|---|---|---|---|---|---|---|---|---|
| event year -2 | 0.019** | 0.002 | -0.011** | -0.000 | 0.000 | -0.006 | -0.007 | -0.001 | 0.003 | -0.015 | 0.021 | -0.045 | -0.006 |
| | (0.008) | (0.006) | (0.005) | (0.011) | (0.007) | (0.007) | (0.037) | (0.006) | (0.010) | (0.011) | (0.031) | (0.074) | (0.042) |
| event year 0 | 0.031*** | 0.039*** | 0.037*** | 0.048*** | 0.025*** | 0.038*** | -0.009 | 0.026*** | 0.017 | 0.049*** | 0.104*** | 0.041 | -0.030 |
| | (0.011) | (0.006) | (0.005) | (0.012) | (0.008) | (0.007) | (0.040) | (0.007) | (0.014) | (0.013) | (0.032) | (0.048) | (0.073) |
| event year 1 | 0.058*** | 0.057*** | 0.057*** | 0.061*** | 0.028*** | 0.065*** | 0.067** | 0.046*** | 0.028* | 0.070*** | 0.176*** | 0.466 | 0.067 |
| | (0.013) | (0.007) | (0.006) | (0.014) | (0.009) | (0.008) | (0.031) | (0.008) | (0.016) | (0.013) | (0.040) | (0.400) | (0.077) |
| $DD_{idst}$ x event year -2 | -0.019 | 0.003 | 0.014** | 0.015 | 0.006 | 0.003 | 0.022 | 0.003 | 0.005 | 0.021 | -0.004 | 0.049 | -0.008 |
| | (0.012) | (0.007) | (0.007) | (0.014) | (0.009) | (0.009) | (0.047) | (0.009) | (0.016) | (0.015) | (0.039) | (0.147) | (0.114) |
| $DD_{idst}$ x event year 0 | -0.066*** | -0.033*** | -0.016** | -0.085*** | -0.017 | -0.068*** | -0.160* | -0.314*** | -0.062** | -0.118*** | -0.185*** | -3.116*** | -0.060 |
| | (0.023) | (0.011) | (0.008) | (0.022) | (0.014) | (0.014) | (0.091) | (0.021) | (0.027) | (0.027) | (0.057) | (1.081) | (0.134) |
| $DD_{idst}$ x event year 1 | -0.075*** | -0.061*** | -0.028*** | -0.069*** | -0.031** | -0.055*** | -0.408*** | -0.163*** | -0.089*** | -0.045** | -0.213*** | -1.728** | -0.101 |
| | (0.024) | (0.012) | (0.010) | (0.024) | (0.016) | (0.013) | (0.107) | (0.016) | (0.029) | (0.022) | (0.061) | (0.840) | (0.138) |
| Constant | 13.218*** | 13.194*** | 13.307*** | 13.109*** | 13.107*** | 13.273*** | 12.919*** | 13.169*** | 13.127*** | 13.272*** | 12.917*** | 12.426*** | 13.059*** |
| | (0.005) | (0.002) | (0.002) | (0.005) | (0.003) | (0.003) | (0.017) | (0.003) | (0.005) | (0.005) | (0.012) | (0.157) | (0.027) |
| Number of IDs | 29,642 | 91,623 | 112,493 | 47,846 | 70,343 | 86,172 | 5,744 | 120,563 | 28,492 | 36,062 | 12,708 | 265 | 2,334 |
| R-squared | 0.001 | 0.001 | 0.002 | 0.001 | 0.000 | 0.001 | 0.009 | 0.009 | 0.001 | 0.002 | 0.003 | 0.186 | 0.001 |
| Obs. | 6,060 | 18,738 | 22,778 | 9,782 | 14,427 | 17,582 | 1,186 | 24,499 | 5,851 | 7,365 | 2,621 | 57 | 487 |
| *F-test: $DD_{idst}$ x event year -2 =0* | *0.102* | *0.672* | *0.0342* | *0.298* | *0.498* | *0.762* | *0.636* | *0.749* | *0.758* | *0.164* | *0.928* | *0.741* | *0.945* |
| *Standardised difference* | *0.017* | *0.004* | *0.013* | *0.012* | *0.010* | *0.012* | *0.076* | *0.002* | *0.004* | *0.004* | *0.022* | *0.123* | *0.000* |

*Notes*: Models are estimated according to equation (1) by replacing the combined indicator for a health shock with event years, separately by disease group. Event years -3 and -1 are reference categories. Standard errors clustered at a (experimental) individual level are in parentheses.

*** $p<0.01$, ** $p<0.05$, * $p<0.1$

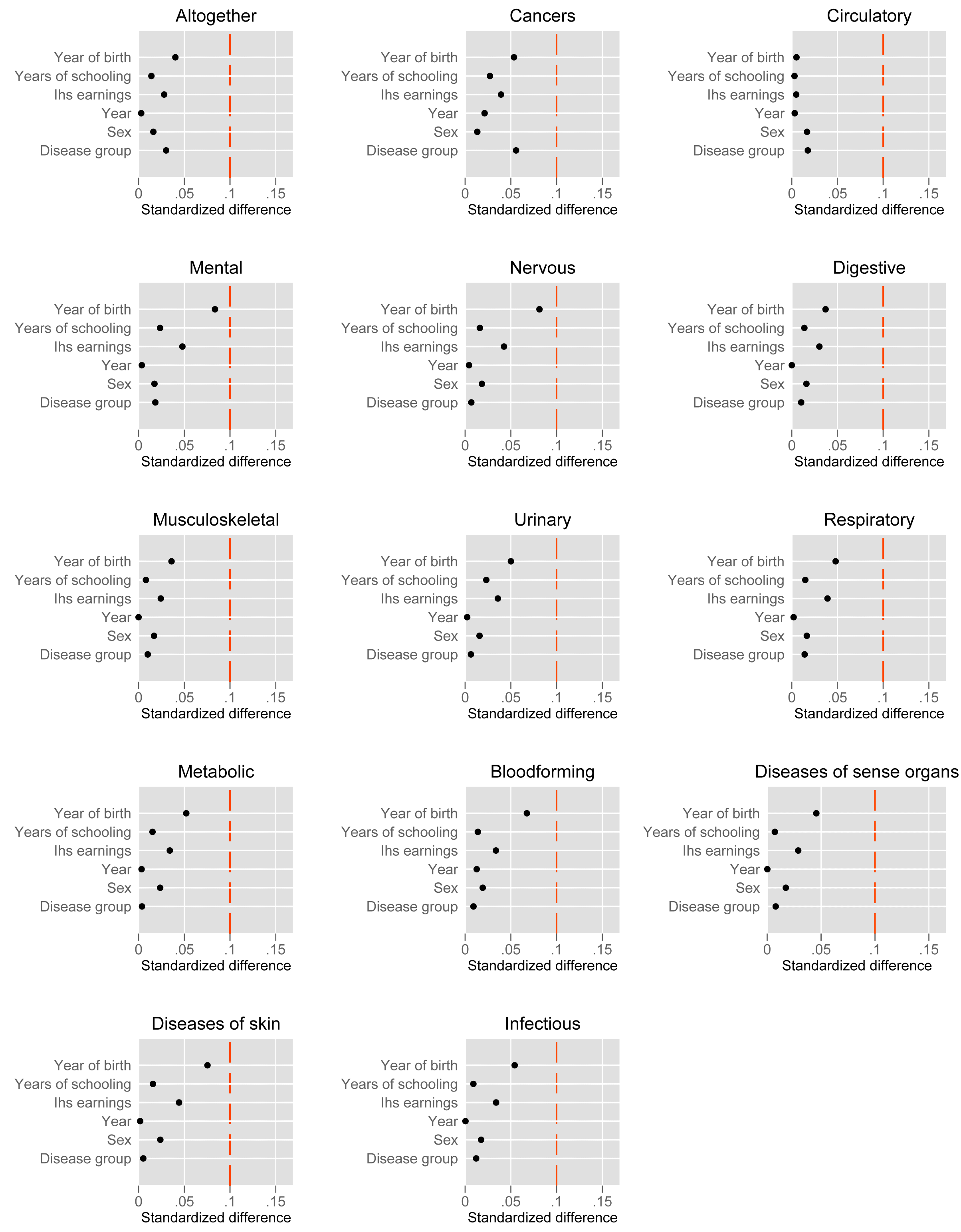

Figure B1 – Results of the balancing test for the estimation sample, in total and by ICD-chapter disease groups

*Notes*: The figures show the results of the test for the covariate balance using absolute standardized differences with a threshold of 0.1, with values below commonly interpreted as indicating balance.

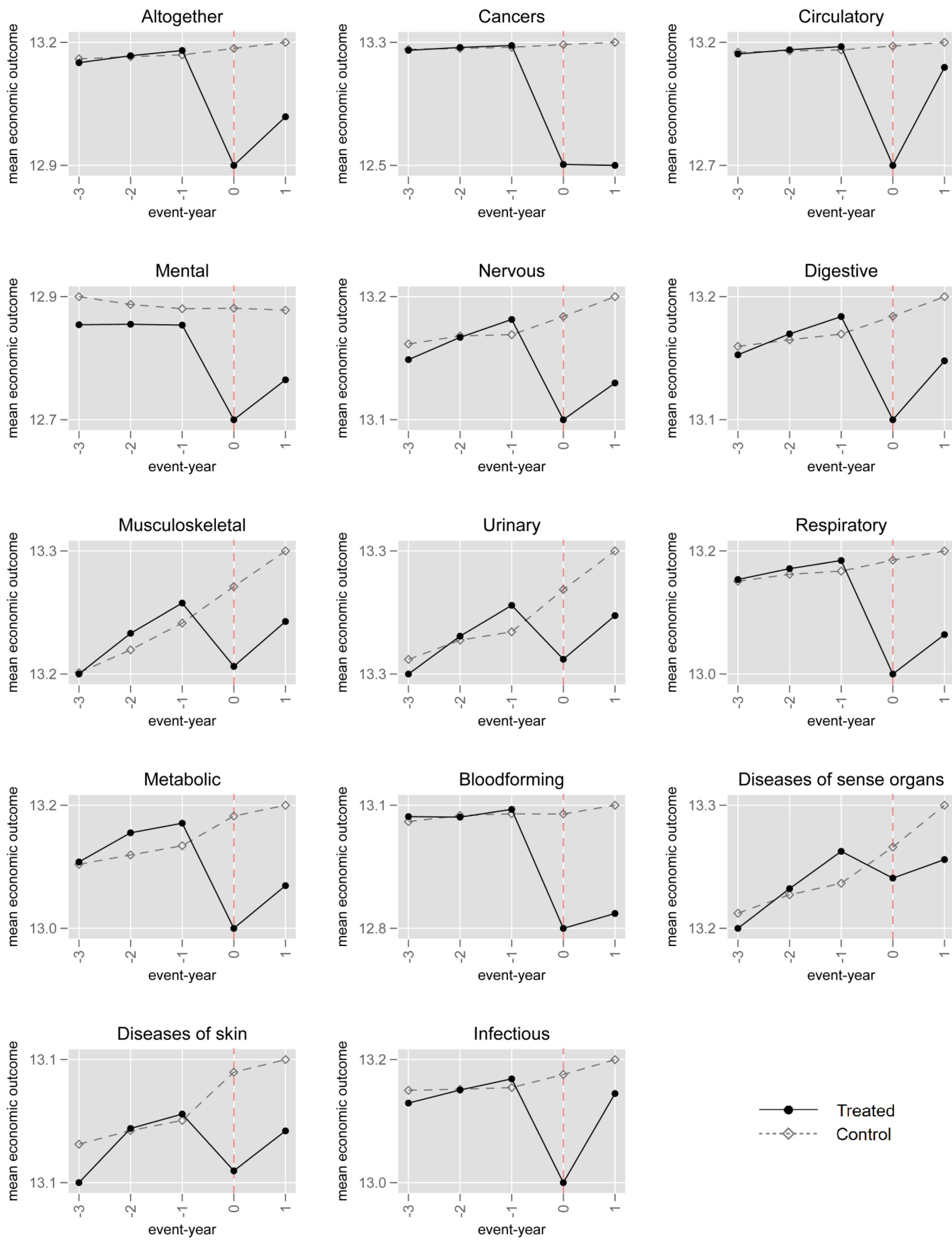

Figure B2 – Development of **family income** (IHS) by event-years for treated and control individuals in the estimation sample, in total and by ICD-chapter disease groups

*Notes*: The figure shows sample means of the outcomes by treatment group and event year dissagregated by ICD chapter.

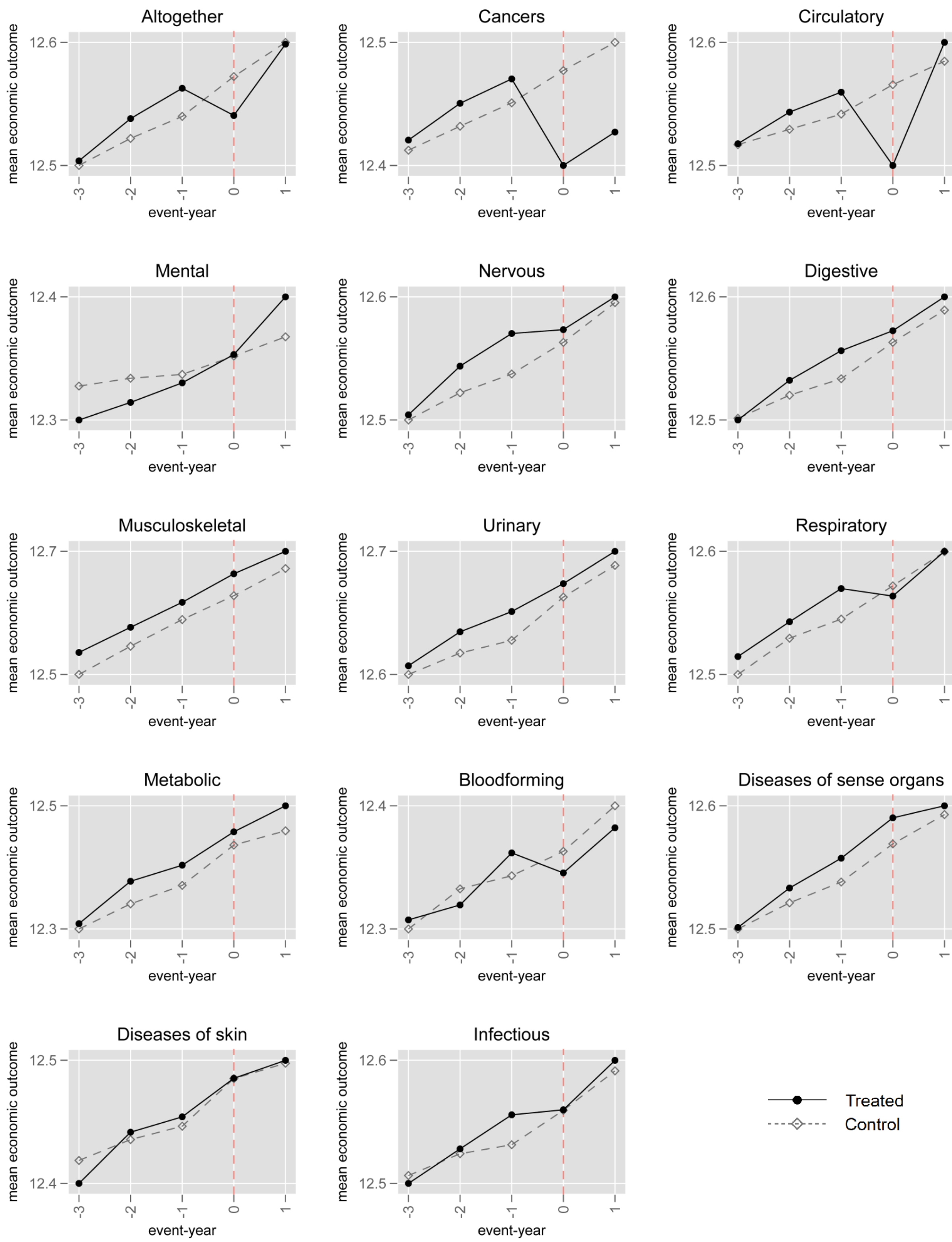

Figure B3 – Development of the **personal income** (IHS) by event-years for treated and control individuals in the estimation sample, in total and by ICD-chapter disease groups

*Notes*: The figure shows sample means of the outcomes by treatment group and event year dissagregated by ICD chapter.

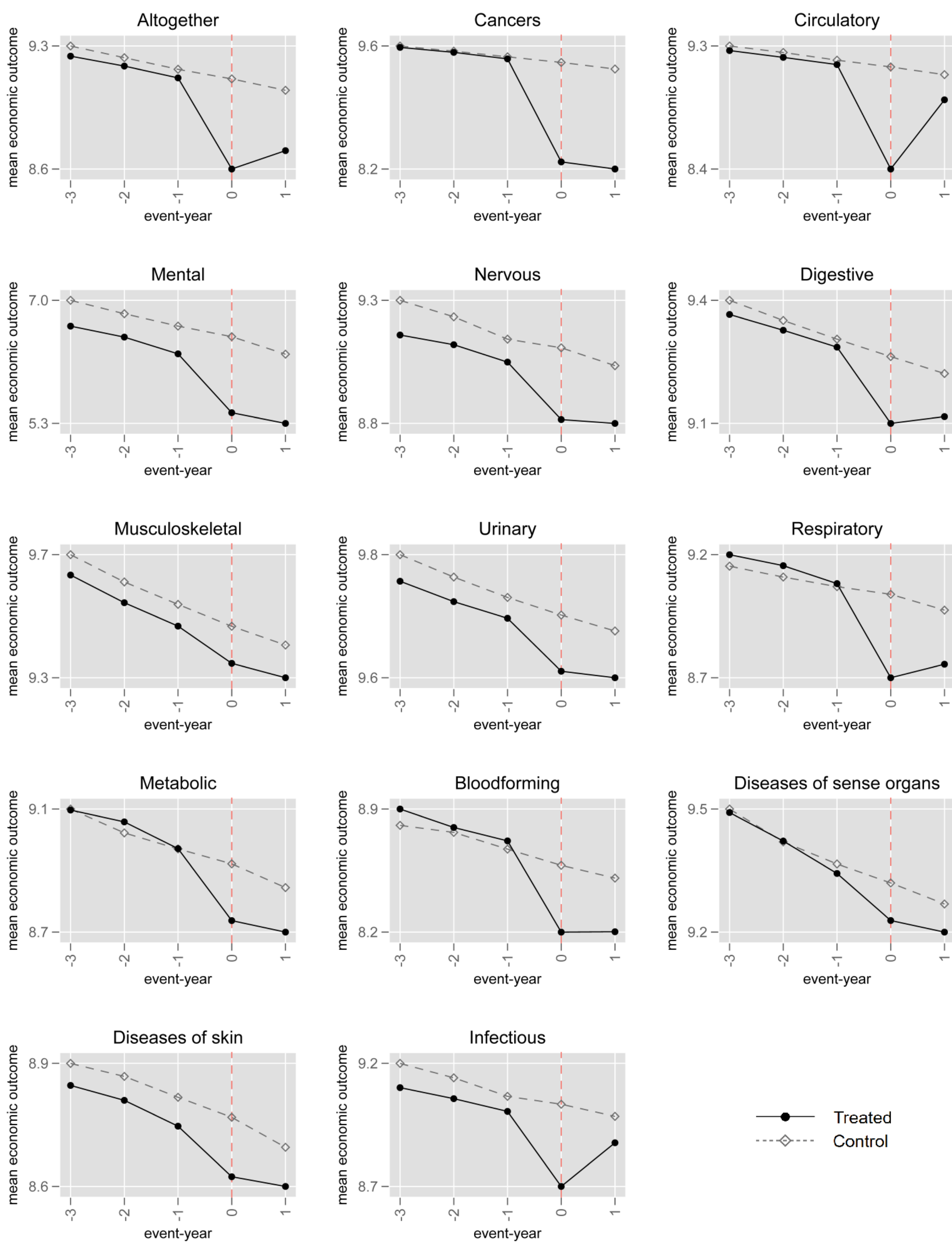

Figure B4 – Development of the **partner income** (IHS) by event-years for treated and control individuals in the estimation sample, in total and by ICD-chapter disease groups

*Notes*: The figure shows sample means of the outcomes by treatment group and event year dissagregated by ICD chapter.

## Appendix C Additional Analyses

Table C1. **Mental Diseases**: Impact of a Health Shock and Mediating Impact of Pharmaceutical Treatments on *Outpatient Hospital Visits* and the Sources *IHS Personal Incomes*: DD and DDD Specifications

| | Income | Outpatient hospital visits | Wage | Sickness benefits | Unempl. benefits | Disability benefits | Early retirement benefits | Capital income |
|---|---|---|---|---|---|---|---|---|
| | (1) | (2) | (3) | (4) | (5) | (6) | (7) | (8) |
| (A) DD: | | | | | | | | |
| $treat_i \times post_{it}$ | 0.0267*** | 1.116*** | -0.578*** | 1.921*** | 0.149*** | 0.353*** | 0.0110 | 0.128*** |
| | (0.008) | (0.027) | (0.021) | (0.026) | (0.007) | (0.015) | (0.028) | (0.029) |
| (B) DDD: | | | | | | | | |
| $treat_i \times post_{it} \times M_{dt}$ | 0.00732*** | -0.111*** | -0.00658 | -0.192*** | -0.00194 | -0.00867 | -0.00486 | -0.0166** |
| | (0.000) | (0.009) | (0.005) | (0.007) | (0.002) | (0.011) | (0.005) | (0.008) |
| Outcome for $treat_i \times post_{it} = 0$ | 20.187 | 0.611 | 17.903 | 1.723 | 0.0665 | 2.475 | 6.146 | 0.423 |
| Observations | 89 737 | 89 737 | 89 737 | 89 737 | 89 737 | 89 737 | 20 240 | 89 737 |
| Number of individuals | 33 851 | 33 851 | 33 851 | 33 851 | 33 851 | 33 851 | 9 305 | 33 851 |

*Notes*: This table reports estimates of the regression coefficient on $treat_i \times post_{it}$ from equation (1) (panel A) and on $treat_i \times post_{it} \times M_{dt}$ from equation (2) (panel B) for a group of mental diseases (ICD-10, chapter F). The sample comprises observations from 2001, the year the outpatient register became available. The sample for the column 5 includes records for the individuals between age 55 and 67 (eligible to benefits). $M_{dt}$ denote pharmaceutical treatments that are a one-year lag of cumulative count of new molecular entities available in year $t$ to treat disease $d$ (see Section II.B for details). For treated individuals, $M_{dt}$ are fixed at the level of disease group and year of hospitalization. Individuals receive separate identifiers when they appear in both the treated and control groups. Outcome for $treat_i \times post_{it} = 0$ is provided for the absolute outcome, in units for column 2 and in 10,000 SEK in columns 1 and 3−8. Standard errors, clustered at the individual (experimental) level, are reported in parentheses.

*** $p < 0.01$, ** $p < 0.05$, * $p < 0.10$

Table C2. **Distributional Impacts**: Mediating Impact of Pharmaceutical Treatments across the Quartiles of Pre-Shock Income Levels on the Sources of IHS *Family*, *Personal, and Partner Incomes,* DDD Specifications

| | Family Income | Personal Income | Partner Income |
|---|---|---|---|
| | (1) | (2) | (3) |
| **First Quartile (Lowest)** | | | |
| (B) DDD: | | | |
| $treat_i \times post_{it} \times M_{dt}$ | 0.0188*** | 0.00740*** | 0.0238*** |
| | (0.000) | (0.000) | (0.000) |
| **Second Quartile** | | | |
| (B) DDD: | | | |
| $treat_i \times post_{it} \times M_{dt}$ | 0.0180*** | 0.00696*** | 0.0230*** |
| | (0.000) | (0.000) | (0.000) |
| **Third Quartile** | | | |
| (B) DDD: | | | |
| $treat_i \times post_{it} \times M_{dt}$ | 0.0157*** | 0.00527*** | 0.0222*** |
| | (0.000) | (0.000) | (0.000) |
| **Fourth Quartile (Highest)** | | | |
| (B) DDD: | | | |
| $treat_i \times post_{it} \times M_{dt}$ | 0.0129*** | 0.00154*** | 0.0127*** |
| | (0.000) | (0.000) | (0.000) |
| Outcome for $treat_i \times post_{it} = 0$, 10 000 SEK | 32.803 | 18.909 | 13.894 |
| Observations | 11 032 884 | 11 032 884 | 11 032 884 |
| Number of individuals | 2 243 040 | 2 243 040 | 2 243 040 |

*Notes*: This table reports estimates of the regression coefficient on $treat_i \times post_{it} \times M_{dt}$ from equation (2) across quartiles of pre-shock IHS income (family, personal, or partner income, respectively). $M_{dt}$ denote pharmaceutical preatments that are a one-year lag of cumulative count of new molecular entities available in year *t* to treat disease *d*. For treated individuals, $M_{dt}$ are fixed at the level of disease group and year of hospitalization. Individuals receive separate identifiers when they appear in both the treated and control groups. Standard errors, clustered at the individual (experimental) level, are reported in parentheses.

*** $p < 0.01$, ** $p < 0.05$, * $p < 0.10$

## Appendix D Robustness Analyses

Table C1. **Compositional Changes**: Impact of a Health Shock and Mediating Impact of Pharmaceutical Treatments on the Sources of IHS *Family*, *Personal, and Partner Incomes,* DD and DDD Specifications

| | Family Income | Personal Income | Partner Income |
|---|---|---|---|
| | (1) | (2) | (3) |
| **I Balanced Sample** | | | |
| (A) DD: | | | |
| $treat_i \times post_{it}$ | -0.310*** | -0.047*** | -0.459*** |
| | (0.002) | (0.002) | (0.004) |
| (B) DDD: | | | |
| $treat_i \times post_{it} \times M_{dt}$ | 0.0166*** | 0.00483*** | 0.0214*** |
| | (0.000) | (0.000) | (0.000) |
| Outcome for $treat_i \times post_{it} = 0$, 10 000 SEK | 32.942 | 18.982 | 13.960 |
| Observations | 10 547 440 | 10 547 440 | 10 547 440 |
| Number of individuals | 2 109 488 | 2 109 488 | 2 109 488 |
| **II Reweighted with Inverse Probability Weights** | | | |
| (A) DD: | | | |
| $treat_i \times post_{it}$ | -0.304*** | -0.053*** | -0.449*** |
| | (0.002) | (0.002) | (0.004) |
| (B) DDD: | | | |
| $treat_i \times post_{it} \times M_{dt}$ | 0.0163*** | 0.00520*** | 0.0207*** |
| | (0.000) | (0.000) | (0.000) |
| Outcome for $treat_i \times post_{it} = 0$, 10 000 SEK | 33.260 | 18.885 | 14.374 |
| Observations | 10 093 284 | 10 093 284 | 10 093 284 |
| Number of individuals | 2 052 772 | 2 052 772 | 2 052 772 |
| **III Adding Event-Year X ICD Chapter Fixed Effects** | | | |
| (A) DD: | | | |
| $treat_i \times post_{it}$ | -0.313*** | -0.050*** | -0.463*** |
| | (0.002) | (0.002) | (0.004) |
| (B) DDD: | | | |
| $treat_i \times post_{it} \times M_{dt}$ | 0.0165*** | 0.00485*** | 0.0212*** |
| | (0.000) | (0.000) | (0.000) |
| Outcome for $treat_i \times post_{it} = 0$, 10 000 SEK | 32.803 | 18.909 | 13.894 |
| Observations | 11 032 884 | 11 032 884 | 11 032 884 |
| Number of individuals | 2 243 040 | 2 243 040 | 2 243 040 |

*Notes*: This table reports estimates of the regression coefficient on $treat_i \times post_{it}$ from equation (1) (panel A) and on $treat_i \times post_{it} \times M_{dt}$ from equation (2) (panel B). Panel I **Balanced Sample** includes individuals who are observed for all 5 years. Panel II **Individuals with Inverse Probability Weights** sample is the same as in Table 1. Inverse probability weights are obtained from a probit regression of whether an individual is observed in only one event-year after the shock on whether the individual had financial debts in the year prior to the shock. Panel III **Adding Event-Year X ICD Chapter Fixed Effects** is estimated on the same sample as for Table 1. $M_{dt}$ denote pharmaceutical preatments that are a one-year lag of cumulative count of new molecular entities available in year *t* to treat disease *d* (see Section II.B for details). For treated individuals, $M_{dt}$ are fixed at the level of disease group and year of hospitalization. Individuals receive separate identifiers when they appear in both the treated and control groups. Standard errors, clustered at the individual (experimental) level, are reported in parentheses.

*** p < 0.01, ** p < 0.05, * p < 0.10

Table C2. **Alternative Medical Innovation Measures**: Impact of a Health Shock and Mediating Impact of Pharmaceutical Treatments on the Sources of IHS *Family*, *Personal, and Partner Incomes,* DD and DDD Specifications

| | Family Income | Personal Income | Partner Income |
|---|---|---|---|
| | (1) | (2) | (3) |
| **I 1-Year Lags of Patents** | | | |
| DDD: | | | |
| $treat_i \times post_{it} \times M_{dt}$ | 0.0350*** | 0.0103*** | 0.0409*** |
| | (0.000) | (0.000) | (0.000) |
| **II 5-Year Lags of Pharmaceutical Treatments** | | | |
| DDD: | | | |
| $treat_i \times post_{it} \times M_{dt}$ | 0.0187*** | 0.00497*** | 0.0247*** |
| | (0.000) | (0.000) | (0.000) |
| **III First Differences of 1-Year Lags of Pharmaceutical Treatments** | | | |
| DDD: | | | |
| $treat_i \times post_{it} \times M_{dt}$ | 0.0209*** | 0.00987*** | 0.0284*** |
| | (0.000) | (0.000) | (0.000) |
| **IV Non-Swedish Pharmaceutical Treatments** | | | |
| DDD: | | | |
| $treat_i \times post_{it} \times M_{dt}$ | 0.0273*** | 0.00801*** | 0.0349*** |
| | (0.000) | (0.000) | (0.000) |
| **V Without Short-Lived Pharmaceutical Treatments** | | | |
| DDD: | | | |
| $treat_i \times post_{it} \times M_{dt}$ | 0.0167*** | 0.00485*** | 0.0214*** |
| | (0.000) | (0.000) | (0.000) |
| **VI Only Never-Disaproved Pharmaceutical Treatments** | | | |
| DDD: | | | |
| $treat_i \times post_{it} \times M_{dt}$ | 0.0136*** | 0.00390*** | 0.0175*** |
| | (0.000) | (0.000) | (0.000) |
| Outcome for $treat_i \times post_{it} = 0$, 10 000 SEK | 32.803 | 18.982 | 13.894 |
| Observations | 11 032 884 | 10 547 440 | 11 032 884 |
| Number of individuals | 2 243 040 | 2 109 488 | 2 243 040 |

*Notes*: This table reports estimates of the regression coefficient on $treat_i \times post_{it} \times M_{dt}$ from equation (2). Panel I **1-Year Lags of Patents** uses one-year lags of patents for diagnostics, therapy, and surgery (in tens) as a measure of medical innovation $M_{dt}$ (its mean is 1.6 and a standard deviation is 2.43). Panel II **5-year Lags of Pharmaceutical Treatments** uses 5-year lags of pharmaceutical treatments as a measure of medical innovation $M_{dt}$. Panel III **First Differences of 1-Year Lags of Pharmaceutical Treatments** uses first differences of pharmaceutical treatments (1-year lags of) instead of levels of pharmaceutical treatments (1-year lags of). Panel IV **Non-Swedish Pharmaceutical Treatments** includes only (1-year lags of) pharmaceutical treatments with a non-Swedish origin. Panel V **Without Short-Lived Pharmaceutical Treatments** excludes from pharmaceutical treatments those that were disapproved within a 5-year period since approval (70 pharmaceutical treatments). Panel VI **Only Never-Disaproved Pharmaceutical Treatments** relies on pharmaceutical treatments that remained approved across the sample period (1548 pharmaceutical treatments). For treated individuals, $M_{dt}$ are fixed at the level of disease group and year of hospitalization. Individuals receive separate identifiers when they appear in both the treated and control groups. Standard errors, clustered at the individual (experimental) level, are reported in parentheses.

*** $p < 0.01$, ** $p < 0.05$, * $p < 0.10$

Table C3. **An Alternative Estimator**: Impact of a Health Shock and Mediating Impact of Pharmaceutical Treatments on the Sources of IHS *Family*, *Personal, and Partner Incomes,* DD and DDD Specifications

| | Family Income | Personal Income | Partner Income |
|---|---|---|---|
| | (1) | (2) | (3) |
| **Callaway and Sant'Anna (2021)'s estimator** | | | |
| (A) DD for the Q1 (lowest) | | | |
| $treat_i \times post_{it}$ | -0.401*** | -0.155*** | -0.524*** |
| | (0.032) | (0.057) | (0.092) |
| (B) DDD for the Q4 (highest) | -0.098*** | -0.0193 | -0.256*** |
| $treat_i \times post_{it}$ | (0.007) | (0.010) | (0.022) |
| | | | |
| Outcome for $treat_i \times post_{it} = 0$, 10 000 SEK | 32.803 | 18.982 | 13.894 |
| Observations | 11 032 884 | 10 547 440 | 11 032 884 |
| Number of individuals | 2 243 040 | 2 109 488 | 2 243 040 |

*Notes*: This table reports estimates of the regression coefficient on $treat_i \times post_{it}$ from equation (1) (panel A) and on $treat_i \times post_{it} \times M_{dt}$ from equation (2) (panel B). **Panel Callaway and Sant'Anna (2021)'s estimator** estimates the effects at the lowest quartile (0–3 pharmaceutical treatments) and highest quartile (15–30 pharmaceutical treatments) of pharmaceutical treatments using estimator proposed by Callaway and Sant'Anna (2021). For treated individuals, $M_{dt}$ are fixed at the level of disease group and year of hospitalization. Individuals receive separate identifiers when they appear in both the treated and control groups. Standard errors, clustered at the individual (experimental) level, are reported in parentheses.

*** $p < 0.01$, ** $p < 0.05$, * $p < 0.10$